\documentclass[11pt]{article}
\usepackage{qcircuit}

\usepackage{url}

\usepackage{hyperref}

\usepackage{amssymb,amscd,amstext,amsmath, amsthm, graphicx}
\usepackage{latexsym}

\usepackage{floatflt}
\usepackage[mathscr]{eucal}
\usepackage{tikz}
\usetikzlibrary{arrows}
\usepackage{verbatim}
\usepackage{soul}
\usepackage{wrapfig}
\usepackage{fancyhdr}
\usepackage{hyperref}
\usepackage{setspace}
\usepackage{mathdots}
\usepackage{blindtext}
\usepackage{enumitem}
 \usepackage{placeins}
 \usepackage{algorithm}
 \usepackage{csquotes}
\usepackage[top=1in, bottom=1in, left=1in, right=1in]{geometry}

\newcommand{\ket}[1]{\ensuremath{\left|#1\right\rangle}}

\definecolor{linkcolor}{HTML}{000000}
\definecolor{urlcolor}{HTML}{000000}
\hypersetup{pdfstartview=Fit, linkcolor=linkcolor,urlcolor=urlcolor,
colorlinks=true, citecolor=linkcolor}
\renewenvironment{abstract}{%
\begin{center}\begin{minipage}{0.9\textwidth}\begin{small}
\textbf{Abstract.}}
{\end{small}\par\noindent\end{minipage}\end{center}}

\title{\bf The Fifth International Students' Olympiad in Cryptography --- NSUCRYPTO: \\problems and their solutions\footnote{The paper was supported by the Russian Ministry of Science and Education (the 5-100 Excellence Programme and the Project no.\,1.13559.2019/13.1), by the Russian Foundation for Basic Research (projects no.\,18-07-01394, 18-31-00479, 18-31-00374), by the program of fundamental scientific researches of the SB RAS no.\,I.5.1, project no.\,0314-2019-0017.}
}

\author{A.~Gorodilova$^{1,2}$,
    S.~Agievich$^{3}$,
    C.~Carlet$^{4}$,
    X.~Hou$^{5}$,
    V.~Idrisova$^{1,2}$,\\
    N.~Kolomeec$^{1,2}$,
    A.~Kutsenko$^{1}$,
    L.~Mariot$^{6}$
    A.~Oblaukhov$^{1,2}$,
    S.~Picek$^{7}$,\\
    B.~Preneel$^{8}$,
    R.~Rosie$^{9}$,
    N.~Tokareva$^{1,2,10}$
    \\
  \\
  {\small$^{1}$Novosibirsk State University, Novosibirsk, Russia}\\
  {\small$^{2}$Sobolev Institute of Mathematics, Novosibirsk, Russia} \\
  {\small$^{3}$Belarusian State University, Minsk, Belarus} \\
  {\small$^{4}$University of Paris 8, Paris, France} \\
  {\small$^{5}$University of South Florida, Tampa, FL 33620, United States of America} \\
  {\small$^{6}$ DISCo, Universit`a degli Studi di Milano-Bicocca, Viale Sarca 336/14, 20126 Milano, Italy}\\
  {\small$^{7}$Cyber Security Research Group, Delft University of Technology, Mekelweg 2, Delft, The Netherlands} \\  %Delft University of Technology, Delft, The Netherlands,} \\
  {\small$^{8}$ESAT-COSIC, KU Leuven, Leuven, Belgium} \\
  {\small$^{9}$ University of Luxembourg, Luxembourg, Luxembourg}\\
  {\small$^{10}$ JetBrains Research, Novosibirsk, Russia}\\
    \\
    {\small E-mail: {\tt nsucrypto@nsu.ru}}
    }

\date{}

\begin{document}

\hypersetup{pageanchor=false}

\begin{titlepage}
\maketitle
\begin{abstract}
Problems and their solutions of the Fifth International Students' Olympiad in cryptography NSUCRYPTO'2018 are presented. We consider problems related to attacks on ciphers and hash functions, Boolean functions, quantum circuits, Enigma, etc. We discuss several open problems on orthogonal arrays, Sylvester matrices and disjunct matrices. The problem of existing an invertible Sylvester matrix whose inverse is again a Sylvester matrix was completely solved during the Olympiad.

\vspace{0.2cm}

\end{abstract}
\end{titlepage}

\hypersetup{pageanchor=true}
\pagenumbering{arabic}

\section*{Introduction}

NSUCRYPTO --- The International Students' Olympiad in cryptography --- celebrated its 5-year anniversary in 2018. Interest in the Olympiad around the world is significant: there were more than 1600 participants from 52 countries in the first five Olympiads from 2014 to 2018! The Olympiad program committee includes specialists from Belgium, France, The Netherlands, USA, Norway, India, Belarus', and Russia.

Let us shortly formulate the format of the Olympiad. One of the Olympiad main ideas is that everyone can participate! Each participant chooses his/her category when registering on the Olympiad website \href{https://nsucrypto.nsu.ru/}{\textcolor{blue}{\tt nsucrypto.nsu.ru}}. There are three categories: ``school students'' (for junior researchers: pupils and high school students), ``university students'' (for participants who are currently studying at universities) and ``professionals'' (for participants who have already completed education or just want to be in the restriction-free category). Awarding of the winners is held in each category separately.

The Olympiad consists of two independent Internet rounds: the first one is individual (duration 4 hours 30 minutes) while the second round is team (duration 1 week). The first round is divided into two sections: A --- for ``school students'', B --- for ``university students'' and ``professionals''. The second round is general for all participants. Participants read the Olympiad problems and submit their solutions using the Olympiad website. The language of the Olympiad is English.

The Olympiad participants are always interested in solving different problems of various complexities at the intersection of mathematics and cryptography. They show their knowledge, creativity and professionalism. That is why the Olympiad not only includes interesting tasks with known solutions but also offers unsolved problems in this area. This year, one of such open problems, ``Sylvester matrices'', was completely solved by three teams! All the open problems stated during the Olympiad history can be found at
\href{https://nsucrypto.nsu.ru/unsolved-problems}{\textcolor{blue}{\tt nsucrypto.nsu.ru/unsolved-problems}}. On the website we also mark the current status of each problem. For example, in addition to ``Sylvester matrices'' solved in 2018, the problem ``Algebraic immunity'' was completely solved during the Olympiad in 2016. And what is important for us, some participants were trying to find solutions after the Olympiad was over. For example, a partial solution for the problem ``A secret sharing'' (2014) was proposed in \cite{sharing}.
We invite everybody who has ideas on how to solve the problems to send your solutions to~us!

The paper is organized as follows. We start with problem structure of the Olympiad in section~\ref{problem-structure}. Then we present formulations of all the problems stated during the Olympiad and give their detailed solutions in section~\ref{problems}. Finally, we publish the lists of NSUCRYPTO'2018 winners in section \ref{winners}.

Mathematical problems of the previous International Olympiads NSUCRYPTO'2014, NSUCRYPTO'2015, NSUCRYPTO'2016, and NSUCRYPTO'2017 can be found in \cite{nsucrypto-2014}, \cite{nsucrypto-2015}, \cite{nsucrypto-2016}, and \cite{nsucrypto-2017} respectively.

\section{Problem structure of the Olympiad}
\label{problem-structure}

There were 16 problems stated during the Olympiad, some of them were included in both rounds (Tables\;\ref{Probl-First},\,\ref{Probl-Second}). Section A of the first round consisted of six problems, whereas the section B contained seven problems. Three problems were common for both sections. The second round was composed of eleven problems. Three problems of the second round were marked as unsolved (awarded special prizes from the Program Committee).

\begin{table}[ht]
\centering\footnotesize
\caption{Problems of the first round}
\medskip
\label{Probl-First}
\begin{tabular}{cc}
\begin{tabular}{|c|l|c|}
  \hline
  N & Problem title & Maximum scores \\
  \hline
  \hline
  1 & A digital signature & 4 \\
    \hline
  2 & Jack and the Beanstalk & 4 \\
    \hline
  3 & Key matrices & 4 \\
    \hline
  4 & A sequence & 4 \\
    \hline
  5 & Solutions of the equation & 4 \\
    \hline
  6 & Stickers & 6 \\
  \hline
\end{tabular}
&
\begin{tabular}{|c|l|c|}
  \hline
  N & Problem title & Maximum scores \\
  \hline
    \hline
  1 & Stickers & 6 \\
    \hline
  2 & Key matrices & 4 \\
    \hline
  3 & A sequence & 4 \\
    \hline
  4 & Quantum circuits & 4 \\
    \hline
  5 & {\tt Bash-S3} & 8 \\
    \hline
  6 & Metrical cryptosystem --- 2 & 6 \\
  \hline
  7 & A fixed element & 10 \\
  \hline
\end{tabular}
\\
\noalign{\smallskip}
Section A
&
Section B
\\
\end{tabular}
\end{table}

\begin{table}[ht]
\centering\footnotesize
\caption{Problems of the second round}
\smallskip
\label{Probl-Second}
\begin{tabular}{|c|l|c|}
  \hline
  N & Problem title & Maximum scores \\
  \hline
    \hline
  1 & A digital signature & 4 \\
    \hline
  2 & Orthogonal arrays & Unsolved \\
    \hline
  3 & Hash function {\tt FNV-1a} & 8 \\
    \hline
  4 & {\tt TwinPeaks2} & 6 \\
    \hline
  5 & An Enigmatic Challenge & 8 \\
    \hline
  6 & Sylvester matrices & Unsolved  \\
    \hline
  7 & Stickers & 6 \\
    \hline
  8 & {\tt Bash-S3} & 8 \\
    \hline
  9 & Metrical cryptosystem --- 2 & 6 \\
    \hline
  10 & A fixed element & 10 \\
    \hline
  11 & Disjunct Matrices & Unsolved \\
  \hline
\end{tabular}
\end{table}

\section{Problems and their solutions}\label{problems}

In this section we formulate all the problems of NSUCRYPTO'2018 and present their detailed solutions paying attention to solutions proposed by the participants.

\subsection{Problem ``A digital signature''}

\subsubsection{Formulation}

Alice uses a new digital signature algorithm, that turns a text message $M$ into a pair $(M,s)$, where~$s$ is an integer and generated in the following way:
\begin{itemize}[noitemsep]
\item[{\bf 1.}] The special function $h$ transforms $M$ into a big positive integer $r=h(M)$.

\item[{\bf 2.}] The number $t=r^2$ is calculated, where $t=\overline{t_1 t_2 \ldots t_n}$.

 \item[{\bf 3.}] The signature $s$ is calculated as $s=t_1+t_2+\ldots+t_n$.
\end{itemize}

Bob obtained the signed message
\begin{center}(\texttt{Congratulations on the fifth year anniversary of NSUCRYPTO!}, 2018)\end{center} from Alice and immediately recognized that something was wrong with the signature! How did he discover it?

\textbf{Remarks}. By $t = \overline{t_1t_2\ldots t_n}$ we mean that $t_1, t_2, \ldots,t_n$ are decimal digits and all digits under the bar form decimal number $t$.

\subsubsection{Solution}

It is widely known that every integer is congruent to the sum of its digits modulo 3. So, we have that $t \equiv_3  2018 \equiv_3 2.$ But $t$ is equal to $r^2$ and a square can not be equal to 2 modulo 3. Thus, we have a contradiction.

We got a lot correct solutions. The most accurate and detailed solutions were sent by Ruxandra Icleanu (Tudor Vianu National College of Computer Science, Romania), Petr Ionov (Yaroslavl State University, Russia), and the team of Henning Seidler and Katja Stumpp (TU Berlin, Germany).

\subsection{Problem ``Jack and the Beanstalk''}
\subsubsection{Formulation}

Little Jack is only seven years old and likes solving riddles involving the powers of two. Recently, his uncle Bitoshi gave him 16 BeanCoin seeds and promised that Jack can collect all BeanCoins which will grow from these seeds. But in order for BeanCoins to grow big and fruitful, Jack must plant the seeds in the garden in a special way. He has to draw eight lines on the ground and plant all 16 seeds on these lines in such a way that each of the lines contains exactly four seeds.

Can you help Jack to achieve his goal and suggest how to plant the seeds?

\subsubsection{Solution}

\begin{figure}[h]
\centering
\begin{tabular}{ccc}
\includegraphics[width=3.0cm]{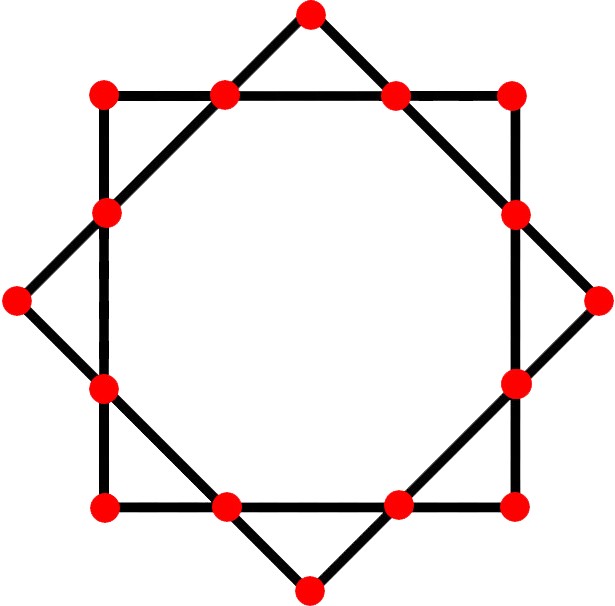}
&
~~~~
&
\includegraphics[width=7.0cm]{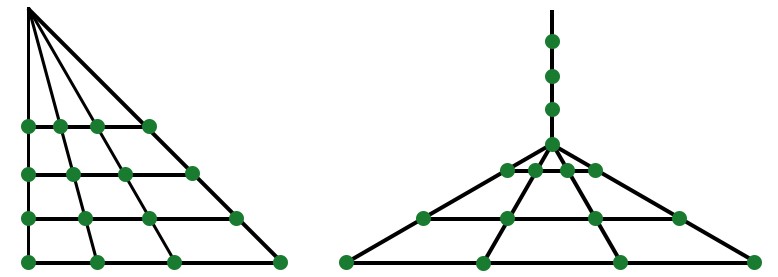}\\
a) octagram
& &
b) solutions by the participants
\end{tabular}
\caption{Lines and seeds}
\label{fig:octagram-solution}
\end{figure}

The seeds can be place on the corners and intersection points of an \textbf{octagram},
as depicted in Figure~\ref{fig:octagram-solution} (a). As is clear from this figure, all eight lines contain exactly four seeds and it is impossible to draw other line contained exactly four seeds.

Many school students found interesting ways to draw these lines, for example Figure~\ref{fig:octagram-solution}~(b). The most interesting ones were given by Gorazd Dimitrov (Yahya Kemal College, Macedonia), Artem Ismagilov (The Specialized Educational and Scientific Center UrFU, Russia), and Igor Pastushenko (The Specialized Educational Scientific Center of Novosibirsk State University, Russia).

\subsection{Problem ``Key matrices''}

\subsubsection{Formulation}

Let $n$ be an {\bf odd} positive integer. In some cipher, a key is a binary $n\times n$ matrix
$$A = \left(
\begin{array}{cccc}
a_{1,1} & a_{1,2} & \dots & a_{1,n} \\
a_{2,1} & a_{2,2} & \dots & a_{2,n} \\
\vdots & \vdots & \ddots & \vdots \\
a_{n,1} & a_{n,2} & \dots & a_{n,n} \\
\end{array}
\right),
$$
where $a_{i,j}$ is either 0 or 1, such that each diagonal of any length $1, 2, \ldots, n-1, n$ contains an \textbf{odd} number of 1s.

What is the minimal and the maximal number of 1s that can be placed in a key matrix~$A$?

{\bf Remarks.} For example, for $n=3$, diagonals are the following ten lines:
\begin{center}
\includegraphics[width=0.25\textwidth]{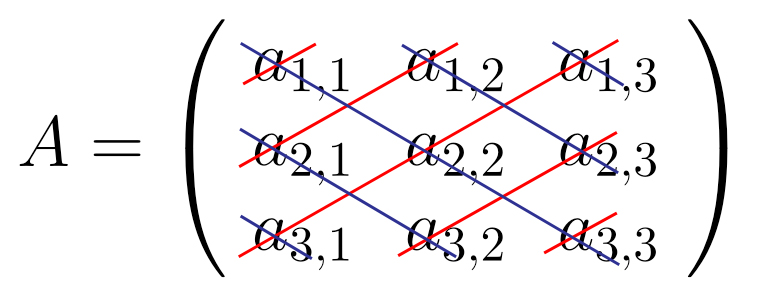}
\end{center}

\subsubsection{Solution}

The correct solution of this problem must consist of two steps. The first step is to find theoretical lower and upper bounds for the number of 1s, and the second step is to prove that these bounds are tight. The best solution was proposed by Aleksei Udovenko (University of Luxembourg), we provide it below.

\textbf{1. Minimum.} Consider the $n \times n$ matrix $A$ ($n$ is odd) with both the top row filled with 1s, the bottom row filled with 1s and the central cell equal to 1; all other elements are 0:

\medskip

$\begin{cases}
a_{1,i}=1, &  1 \leqslant i \leqslant n;
\\ a_{n,i}=1, &  1 \leqslant i \leqslant n;
 \\ a_{(n+1)/2,(n+1)/2}=1 ;
 \\a_{i,j}=0, &  \mbox{otherwise}.
 \end{cases}$
\medskip

 Any diagonal of length less than $n-1$ includes exactly a single 1 (either from the top row or from the bottom row). The two diagonals of length $n$ include three 1s (one from the top row, one from the bottom row and one from the center). Therefore, this matrix satisfies the condition. It has $2n+1$ 1s.

 We now prove that this number of 1s is minimal. Note that each corner cell $a_{1,1}$, $a_{1,n}$, $a_{n,1}$, $a_{n,n}$ makes a single element diagonal. Therefore, these cells must contain 1s. There are $2(n-2)$ diagonals going in the down-right direction and not touching the corners (starting from the cells of the leftmost column and from the cells for the topmost row). Furthermore, the main diagonal without the corner cells must have odd number of 1s too. Therefore, $2n-3$ disjoint diagonals must contain at least one 1, in addition to 4 corner 1s. Therefore, there should be at least $2(n-2) + 1 + 4 = 2n + 1$ 1s in the matrix.

\medskip

\textbf{2. Maximum.} Consider the $n \times n$ matrix $A$ ($n$ is odd) filled with 1s except cells in the leftmost and the rightmost columns which have an even row index:

\medskip

$\begin{cases}
a_{2i,1}=0, &  1 \leqslant i \leqslant (n-1)/2;
\\ a_{2i,n}=0, &  1 \leqslant i \leqslant (n-1)/2;
 \\a_{i,j}=1, &  \mbox{otherwise}.
 \end{cases}$

\medskip

It is easy to check that all diagonals that contain an even number of elements contain a single zero either from the leftmost or from the rightmost column. Therefore, these diagonals have an odd number of 1s. Also, all diagonals that contain an odd number of elements contain no zeroes and thus have an odd number of 1s too. Therefore, this matrix satisfies the condition. It has $n^2-2(n-1)/2 = n^2-n + 1$ 1s.

We now prove that this number is maximal. Consider diagonals going in the down-right direction that have an even number of elements. There are $2(n-1)/2 = (n-1)$ such diagonals and they are disjoint. Each of them must contain at least a single zero. Therefore, the maximum number of 1s is $n^2-n+1$.

\subsection{Problem ``A sequence''}

\subsubsection{Formulation}

Two friends, Roman and Anton, are very interested in sequences and ciphers.
Their new cryptosystem encrypts binary messages of length $n$, $X = (x_1,x_2,\ldots,x_n)$, where each $x_i$ is either 0 or~1. A key $K$ of the cipher is a set of $n$ integers $a_1,a_2,\ldots, a_n$. The ciphertext $Y$ for the message $X$ encrypted with the key $K$ is the integer $$Y = x_1\cdot a_1 + x_2\cdot a_2 + \ldots + x_n\cdot a_n.$$

Roman and Anton change their key regularly. Today, the key $K$ is defined by $$a_i = 2^i + (-1)^i \ \text{ for all } i = 1,\ldots,n.$$

The friends can easily decipher any message using the key defined by this sequence for any $n$!

\begin{enumerate}[noitemsep]
\item[{\bf 1.}] Prove that the encryption is correct for this key $K$ for any $n$: there are no two distinct input messages $X^1$ and $X^2$ such that their ciphertexts $Y^1$ and $Y^2$ are equal, i.\,e. $Y^1 = Y^2$.

\item[{\bf 2.}] Describe an algorithm which can be used to easily decipher any ciphertext $Y$ encrypted with today's key $K$. Here ``easily'' means that the algorithm should work much faster than checking all possible variants for an input message $X$.
\end{enumerate}

\subsubsection{Solution}

Let us firstly show that the sequence $\{a_i\}$ is {\it superincreasing}, i.\,e. $a_{i+1}>\sum_{k=1}^{i}a_k$ for any $i>0$.
Indeed,
$$
\sum_{k=1}^{i}a_k=\sum_{k=1}^{i} (2^k + (-1)^k) = 2^{i+1} - 2 + \sum_{k=1}^i(-1)^k =
\begin{cases}
2^{i+1} - 2, \text{ if }i\text{ is even}\\
2^{i+1} - 3, \text{ if }i\text{ is odd}\\
\end{cases}
<2^{i+1} + (-1)^i = a_{i+1}.
$$

\begin{enumerate}[noitemsep]
\item[{\bf 1.}] Let us show that the encryption is correct. Let $X^1 = (x^1_1, \ldots, x^1_n)$ and $X^2 = (x^2_1, \ldots, x^2_n)$ be two distinct messages, and $i$ is the largest position such that $x^1_i\neq x^2_i$. Without loss of generality, suppose that $x^1_i = 1$. Then
$$
Y^1 - Y^2 =  (x^1_1\cdot a_1 + \ldots + x^1_i\cdot a_i + \ldots + x^1_n\cdot a_n) - (x^2_1\cdot a_1 + \ldots +  x^2_i\cdot a_i + \ldots + x^2_n\cdot a_n)
$$
$$= (x^1_1 - x^2_1) \cdot a_1 + \ldots + (x^1_{i-1} - x^2_{i-1}) \cdot a_{i-1} + a_i > 0
$$
since $\{a_i\}$ is a superincreasing sequence.

\item[{\bf 2.}] The correctness of the decryption algorithm (Algorithm~\ref{alg}) is also based on the superincreasing property of $\{a_i\}$. The complexity of the algorithm consists of $n$ integer comparisons.

\begin{algorithm}
\caption{The decryption algorithm}\label{alg}
    {\bf Input:} $Y$, $n$.

    {\bf Output:} $X = (x_1,\ldots,x_n)$.

    Step 0. $T:=Y$, $i:=n$.

    Step 1. If $T > a_i$, then $x_i = 1$; else $x_i = 0$.

    Step 2. $T:=T - x_i\cdot a_i$, $i:=i-1$. If $i>0$, go to step 1; else return $X$.
\end{algorithm}
\end{enumerate}

\vspace{-0.3cm}
The problem was solved by the majority of participants including eight school students.

\subsection{Problem ``Solutions of the equation''}

\subsubsection{Formulation}

Alice is studying special functions that are used in symmetric ciphers.
Let $E^n$ be the set of all binary vectors $x = (x_1, x_2,\ldots,x_n)$ of length $n$, where $x_i$ is either 0 or~1. Given two vectors $x$ and $y$ from $E^n$ consider their sum $x \oplus y=(x_1\oplus y_1,\ldots, x_n\oplus y_n)$, where $\oplus$ is addition modulo 2.

\textbf{Example}. If $n=3$, then $E^3=\{(000), (001), (010), (011), (100), (101), (110), (111)\}$. Let $x=(010)$ and $y= (011)$, then vector $x \oplus y$ is equal to $(010)\oplus (011)=(0 \oplus 0, 1 \oplus 1, 0 \oplus 1)=(001)$.

We will say that a function $F$ maps $E^n$ to $E^n$ if it transforms any vector $x$ from $E^n$ into some vector $F(x)$ from $E^n$.

\textbf{Example}. Let $n=2$. For instance, we can define $F$ that maps $E^2$ to $E^2$ as follows: $F(00)=(00)$, $F(01)=(10)$, $F(10)=(11)$ and $F(11)=(10)$.

Alice found a function $S$ that maps $E^6$ to $E^6$ in such a way that the vectors $S(x)$ and $S(y)$ are not equal for any nonequal vectors $x$ and $y$. Also, $S$ has another curious property:
the equation
$$S(x) \oplus S(x\oplus a) = b$$ has either 0 or 2 solutions for any nonzero vector $a$ from $E^6$ and any vector $b$ from $E^6$.

Find the number of pairs $(a,b)$ such that this equation has exactly 2 solutions!

\subsubsection{Solution}

Consider a function $S$ that satisfies the conditions of the problem. Let us fix an arbitrary vector $a$ that is nonzero. Consider the set $B_a$ of all possible values of $S(x) \oplus S(x\oplus a)$, i.\,e. $B_a=\{S(x) \oplus S(x\oplus a)~|~x \in E^6\}$. It holds that $|B_a| = 2^5$, since $S(x) \oplus S(x\oplus a)=S(x\oplus a) \oplus S(x\oplus a\oplus a)$. Then for every nonzero $a$ there exist $2^5$ values of $b$, such that $S(x) \oplus S(x\oplus a) = b$ has 2 solutions. Then the number of pairs is equal to $63*32=2016$.

Correct answers were sent by only three school students: Alexey Lvov (Gymnasium 6 of Novosibirsk, Russia), Borislav Kirilov (The First Private Mathematical Gymnasium of Sofia, Bulgaria), and Razvan Andrei Draghici (National College Fratii Buzesti, Romania).

\subsection{Problem ``Quantum circuits''}

\subsubsection{Formulation}

Alice and Bob are interested in quantum circuits. They studied quantum operations and would like to use them for their simple cipher. Let an input plaintext be $P=(p_1,p_2,\ldots,p_{16})\in\mathbb{F}_2^{16}$. The ciphertext $C\in\mathbb{F}_2^{16}$ is calculated as
\begin{equation*}
C=K\oplus \big(F(p_1,\ldots,p_4),\ F(p_5,\ldots,p_8),\ F(p_9,\ldots,p_{12}),\ F(p_{13},\ldots,p_{16})\big),
\end{equation*}
where $K\in\mathbb{F}_2^{16}$ is a secret key and $F$ is a function from $\mathbb{F}_2^4$ to $\mathbb{F}_2^4$; $\oplus$ is bitwise XOR.

The friends found a representation of $F$ from wires and elementary quantum gates which form a quantum circuit. They use Dirac notation and denote computational basis states by $\ket{0}$ and $\ket{1}$. Further, quantum bits (qubits) are considered only in quantum states $\ket{0}$ and $\ket{1}$. Alice and Bob used the following quantum gates and circuit symbols which are given in Table~\ref{circuit-symbols}.

%\vspace{-0.3cm}

\begin{table}[h]
\centering
\caption{Quantum gates and circuit symbols}
\label{circuit-symbols}
\smallskip
{\small
\noindent\begin{tabular}{|p{2.9cm}|p{4.1cm}|p{8.3cm}|}
\hline
Pauli-X gate
&
\hspace{0.6cm}\Qcircuit @C=1.5em @R=1em
	{
		\lstick{\ket{x}} & \gate{X} & \rstick{\ket{x\oplus1}} \qw
	}
&
acts on a single qubit in the state $\ket{x}$, $x\in\{0,1\}$.
\\
\hline
 controlled-NOT gate (CNOT gate)
&
\hspace{0.6cm}\Qcircuit @C=1.5em @R=1em
	{
		\lstick{\ket{x}} & \ctrl{1} & \rstick{\ket{x}} \qw \\
		\lstick{\ket{y}} & \targ & \rstick{\ket{y\oplus x}} \qw
	}
&
acts on two qubits in the states $\ket{x},\ket{y}$, $x,y\in\{0,1\}$;

 it flips the second qubit if and only if the first qubit is in the state~\ket{1}.
\\
\hline
 Toffoli gate\newline (CCNOT gate)
&
\hspace{0.6cm}\Qcircuit @C=1.5em @R=1em
	{
		\lstick{\ket{x}} & \ctrl{1} & \rstick{\ket{x}} \qw \\
		\lstick{\ket{y}} & \ctrl{1} & \rstick{\ket{y}} \qw \\
		\lstick{\ket{z}} & \targ & \rstick{\ket{z\oplus(x\land y)}} \qw
	}
&
acts on three qubits in the states $\ket{x},\ket{y},\ket{z}$, $x,y,z\in\{0,1\}$;
	
it flips the third qubit if and only if the states of the first and the second qubits are both equal to \ket{1}.
\\
\hline
		\hspace{0.6cm}\Qcircuit @C=1.5em @R=1em
		{
			\lstick{\ket{x}} & \meter & \rstick{x} \cw
		}
		& \multicolumn{2}{p{12.4cm}|}{
		a measurement of a qubit in the state $\ket{x}$, $x\in\{0,1\}$, in the computational basis \{\ket{0},\ket{1}\}.}
		\\
		\hline
		\hspace{0.6cm}\Qcircuit @C=2em @R=1em
		{
		&	\qw & \qw
		}
		& \multicolumn{2}{p{12.4cm}|}{
		a wire carrying a single qubit (time goes left to right).}
		\\
		\hline
		\hspace{0.6cm}\Qcircuit @C=2em @R=1em
		{
		&	\cw & \cw
		}
		& \multicolumn{2}{p{12.4cm}|}{
		a wire carrying a single classical bit.}
		\\
		\hline
	\end{tabular}
}
\end{table}

A quantum circuit which describes action of $F$ on $x=\left(x_1,x_2,x_3,x_4\right)\in~\mathbb{F}_2^4$, where $F = (f_1,f_2,f_3,f_4)$ and $f_i,i=1,2,3,4,$ are Boolean functions in $4$ variables, is the following:

\begin{center}
{\footnotesize
~\Qcircuit @C=2em @R=1em
{
	\lstick{\ket{x_1}} & \targ     & \ctrl{1}  & \ctrl{1} & \targ       & \gate{X} & \ctrl{1} & \targ & \ctrl{1} & \targ & \ctrl{1} & \meter & \rstick{f_1(x)}\cw \\
	\lstick{\ket{x_2}} & \ctrl{-1} & \targ     & \ctrl{1} & \ctrl{-1}     & \gate{X} & \targ & \ctrl{-1}         & \targ & \ctrl{-1} & \targ & \meter & \rstick{f_2(x)}\cw\\
	\lstick{\ket{x_3}} & \ctrl{-1} & \ctrl{-1} & \targ    & \targ       & \ctrl{1}      & \targ & \ctrl{-1} & \targ & \gate{X} & \ctrl{1} & \meter & \rstick{f_3(x)}\cw  \\
	\lstick{\ket{x_4}} & \qw       & \gate{X}  & \qw      & \ctrl{-1} & \targ    & \ctrl{-1} & \qw & \ctrl{-1} & \gate{X} & \targ & \meter & \rstick{f_4(x)}\cw
}
}
\end{center}

{\bf The problem.} The friends encrypted the plaintext $P=(0011010111110010)$ and got the ciphertext $C=(1001101010010010)$. Find the secret key $K$!

\subsubsection{Solution}

One can notice that the given circuit can be simplified by observing that the following evolutions
\begin{center}
{\footnotesize
~\Qcircuit @C=2em @R=1em
	{
		\lstick{\ket{x_1}} & \targ     & \ctrl{1}  & \ctrl{1} & \targ       & \gate{X} & \ctrl{1} & \targ & \ctrl{1} & \targ & \ctrl{1} & \meter & \rstick{f_1(x)}\cw \\
		\lstick{\ket{x_2}} & \ctrl{-1} & \targ     & \ctrl{1} & \ctrl{-1}     & \gate{X} & \targ & \ctrl{-1}         & \targ & \ctrl{-1} & \targ & \meter & \rstick{f_2(x)}\cw\\
		\lstick{\ket{x_3}} & \ctrl{-1} & \ctrl{-1} & \targ    & \targ       & \ctrl{1}      & \targ & \ctrl{-1} & \targ & \gate{X} & \ctrl{1} & \meter & \rstick{f_3(x)}\cw  \\
		\lstick{\ket{x_4}} & \qw       & \gate{X}  & \qw      & \ctrl{-1} & \targ    & \ctrl{-1} & \qw & \ctrl{-1} & \gate{X} & \targ & \meter & \rstick{f_4(x)}\cw  \gategroup{3}{5}{4}{7}{.7em}{--}
		\gategroup{1}{9}{2}{11}{.7em}{--}
	}
}
\end{center}

\noindent actually swap two states $\ket{x},\ket{y}$, $x,y\in\{0,1\}$:
\begin{center}
{\footnotesize
~\Qcircuit @C=2em @R=1em
{
	\lstick{\ket{x}} & \ctrl{1} & \targ & \ctrl{1} & \rstick{\ket{y}}\qw \\
	\lstick{\ket{y}} & \targ & \ctrl{-1} & \targ & \rstick{\ket{x}}\qw
}
\hspace{3cm}\Qcircuit @C=2em @R=1em
{
	\lstick{\ket{x}} & \targ & \ctrl{1} & \targ & \rstick{\ket{y}}\qw \\
	\lstick{\ket{y}} & \ctrl{-1} & \targ & \ctrl{-1} & \rstick{\ket{x}}\qw
}
}
\end{center}

\noindent Both of the evolutions

\begin{center}
{\footnotesize
~\Qcircuit @C=2em @R=1em
	{
		\lstick{\ket{x_1}} & \targ     & \ctrl{1}  & \ctrl{1} & \targ       & \gate{X} & \ctrl{1} & \targ & \ctrl{1} & \targ & \ctrl{1} & \meter & \rstick{f_1(x)}\cw \\
		\lstick{\ket{x_2}} & \ctrl{-1} & \targ     & \ctrl{1} & \ctrl{-1}     & \gate{X} & \targ & \ctrl{-1}         & \targ & \ctrl{-1} & \targ & \meter & \rstick{f_2(x)}\cw\\
		\lstick{\ket{x_3}} & \ctrl{-1} & \ctrl{-1} & \targ    & \targ       & \ctrl{1}      & \targ & \ctrl{-1} & \targ & \gate{X} & \ctrl{1} & \meter & \rstick{f_3(x)}\cw  \\
		\lstick{\ket{x_4}} & \qw       & \gate{X}  & \qw      & \ctrl{-1} & \targ    & \ctrl{-1} & \qw & \ctrl{-1} & \gate{X} & \targ & \meter & \rstick{f_4(x)}\cw  \gategroup{1}{5}{2}{7}{.7em}{--}
		\gategroup{3}{9}{4}{11}{.7em}{--}
	}
}
\end{center}

\noindent have form
\begin{center}
{\footnotesize
~\Qcircuit @C=2em @R=1em
	{
		\lstick{\ket{x}} & \targ & \gate{X} & \ctrl{1} & \rstick{\ket{x\oplus y\oplus1}}\qw \\
		\lstick{\ket{y}} & \ctrl{-1} & \gate{X} & \targ & \rstick{\ket{x}}\qw
	}
}
\end{center}
\noindent for $\ket{x},\ket{y}$, $x,y\in\{0,1\}$.

The algebraic normal forms of coordinate Boolean functions of $F$ are
\begin{align*}
f_1(x) &= x_1\oplus x_2x_3, \\
f_2(x) &= x_2\oplus x_1x_4\oplus x_2x_3x_4\oplus1, \\
f_3(x) &= x_3\oplus x_4\oplus x_1x_2\oplus x_1x_3, \\
f_4(x) &= x_4\oplus1,
\end{align*}
where $x\in\mathbb{F}_2^4$. Then
\begin{align*}
K_{1,...,4} &= C_{1,...,4}\oplus F\left(p_1,\ldots,p_4\right)=C_{1,...,4}\oplus(0100)=(1101), \\
K_{5,...,8} &=  C_{5,...,8}\oplus F\left(p_5,\ldots,p_8\right)=C_{5,...,8}\oplus(0010)=(1000), \\
K_{9,...,12} &=  C_{9,...,12}\oplus F\left(p_9,\ldots,p_{12}\right)=C_{9,...,12}\oplus(0000)=(1001), \\
K_{13,...,16} &=  C_{13,...,16}\oplus F\left(p_{13},\ldots,p_{16}\right)=C_{13,...,16}\oplus(0111)=(0101),
\end{align*}
and finally, the key is the following:
\begin{equation*}
K=(1101100010010101).
\end{equation*}

Many participants coped with this problem and correctly found the key.

\subsection{Problem ``Stickers''}

\subsubsection{Formulation}

\begin{figure}[h!]
\centering
\includegraphics[width=0.8\textwidth]{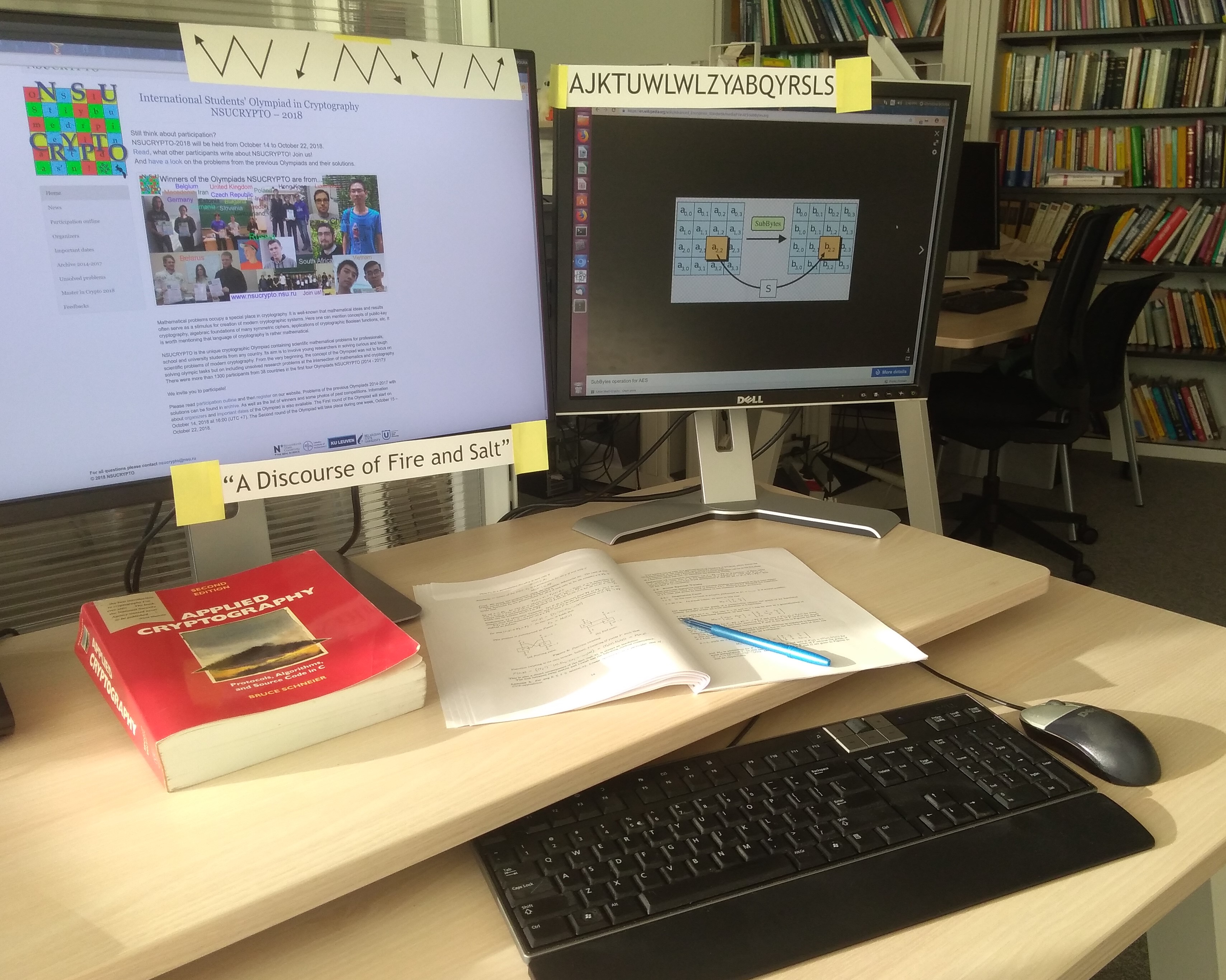}
\vspace{-0.3cm}
\caption{Workspace}
\label{fig:workspace}
\end{figure}

Bob always takes into account all the recommendations of security experts. He
switched from short passwords to long passphrases and changes them every month.
Bob usually chooses passphrases from the books he is reading. Passphrases are
so lengthy and are changed so often! In order to not forget them, Bob decided
to use stickers with hints. He places them on his monitors
(ooh, experts...). The only hope is that Bob's hint system is reliable because
it uses encryption. But is that true? Could you recover Bob's current
passphrase from the photo of his workspace (Figure~\ref{fig:workspace})?

\subsubsection{Solution}

Looking at the picture we see three stickers. One of them is ``A Discourse of Fire and Salt'' that represents a title of a book written by Blaise de Vigen\`{e}re. This is the first hint that probably the Vigin\`{e}re cipher was used.
Then we have a sticker with the ciphertext {\tt AJKTUWLWLZYABQYRSLS} that consists of 19 letters. And finally, we see the sticker with five directed polygonal paths containing a total of 19 vertices. These 19 vertices could correspond to the 19 ciphertext letters.

\begin{figure}[h!]
\centering
\includegraphics[width=7.5cm]{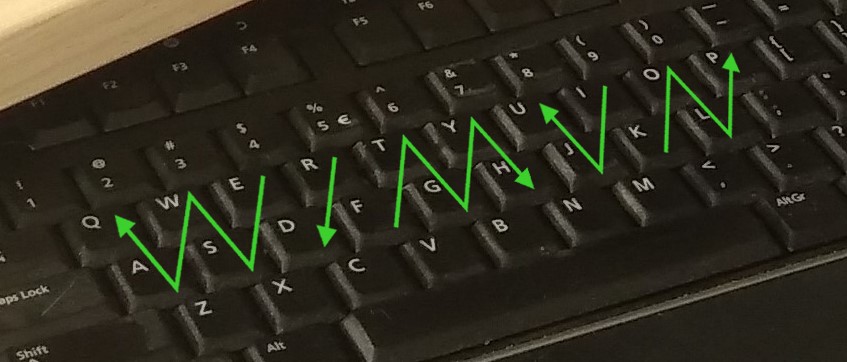}
\vspace{-0.3cm}
\caption{Keyboard rows}
\label{fig:workspace-answer}
\end{figure}

There is a keyboard at the picture. So, we can guess that these arrows could be related to the letters from the keyboard. Let us look at the first two keyboard rows (Figure~\ref{fig:workspace-answer}).
We can recover the secrete key
{\tt ESWAQRDFTGYHIJUKOLP}.
By deciphering the ciphertext using this key and the Vigin\`{e}re cipher, we get {\tt WROTEFIRSTATTHEHEAD}.
Thus, Bob's current passphrase is ``Wrote first at the head''.

Surprisingly, nobody solved this problem in the first round, while five teams solved it in the second round.

\subsection{Problem ``\texttt{Bash-S3}''}

\subsubsection{Formulation}

The sponge function
\href{https://eprint.iacr.org/2016/587}{\textcolor{blue}{\texttt{Bash-f}}} \cite{Agievich-2016}
uses the permu\-tation $S3$ that transforms a triple of 64-bit binary words
$a,b,c$ in the following way:
$$
S3(a,b,c)= (b \vee \neg c\oplus a,\ a \vee c\oplus b,\ a \wedge b\oplus c).
$$
Here $\neg$, $\wedge$, $\vee$, $\oplus$ denote the binary bitwise operations
``NOT', ``AND'', ``OR'', ``XOR'' respectively. The operations are listed in
descending order of priority.
 Let $w^k$ also denote the cyclic shift of
a 64-bit word $w$ to the left by $k\in\{1,2,\ldots,63\}$ positions.

Alice wants to strengthen~$S3$. She can do this by XORing any input~$a,b,c$ or
its cyclic shift to any output. She must use at least one cyclic shift
and she cannot add two identical terms to the same output.

Help Alice change~$S3$ in such a way that a modified $S3$ will still be a
permutation!

{\bf Remarks.}

1. For example, in the expression $b\vee \neg c\oplus a$, we firstly calculate $\neg c$, then calculate $b \vee \neg c$, and after that the final result (according to descending order of operations priority).

2. The modification
$$
(b \vee \neg c\oplus a\oplus a^{11},\
a \vee c\oplus a^7\oplus c,\
a \wedge b\oplus b^{32})
$$
is allowed but it does not satisfy the permutation condition.

3. $S3$ has three outputs: $b \vee \neg c\oplus a,\ a \vee c\oplus b,\ a \wedge b\oplus c$. Alice can add as many inputs and cyclic shifts of inputs as she wants to each of these outputs.
In the remark 2 she add $a^{11}$ to the first output, $b \oplus a^{7} \oplus c$ to the second output, and $c \oplus b^{32}$ to the third output.
Note that the fact that $S3$ is a permutation (as a function $\{0,1\}^{64*3} \to \{0,1\}^{64*3}$) is not obvious. But the problem is only to prove that the modification of $S3$ is a permutation too (as a function $\{0,1\}^{64*3} \to \{0,1\}^{64*3}$).

\subsubsection{Solution}

It is allowed to add to the outputs of $S3$ the outputs of the following linear transformation:
$$
L(a,b,c)=(L_0(a,b,c),L_1(a,b,c),L_2(a,b,c))
$$
that is defined by bitwise XOR operations and cyclic shifts.

The permutation property of a modified $S3$ will be broken if
for some distinct~$(a,b,c)$ and~$(a',b',c')$
\vspace{-0.2cm}
\begin{equation}\label{S3}
S3(a,b,c)\oplus S3(a',b',c')=L(a,b,c)\oplus L(a',b',c').
\end{equation}

We will call the expressions from both sides of equality~\eqref{S3} and the sum  $(a,b,c)\oplus (a',b',c')$ by {\it differences}.
Let
\vspace{-0.2cm}
\begin{align*}
(w_0,w_1,w_2) &= (a,b,c)\oplus (a',b',c'), \\
(W_0,W_1,W_2) &= S3(a,b,c)\oplus S3(a',b',c').
\end{align*}
On the one hand, input and output differences of $S3$ satisfy (for instance, see \cite{Agievich-2016}) the equality
$$
w_0 \wedge W_0 \oplus w_1 \wedge W_1 \oplus w_2 \wedge W_2 = 11\ldots1.
$$
On the other hand, by $(\ref{S3})$ the permutation property of a modified $S3$ will be broken if
$$
	(W_0,W_1,W_2) = L(a,b,c)\oplus L(a',b',c')= L(w_0,w_1,w_2).
$$
As a result, a modified $S3$ will be still a permutation if the following equality %\eqref{Eq.BASH-S3.1}
\begin{equation}\label{Eq.BASH-S3.1}
w_0 \wedge L_0(w_0,w_1,w_2) \oplus
w_1 \wedge L_1(w_0,w_1,w_2) \oplus
w_2 \wedge L_2(w_0,w_1,w_2) = 11\ldots 1
\end{equation}
does not hold for any nonzero $(w_0,w_1,w_2)$. For example, if
$$
L(a,b,c)=(a\oplus a^d\oplus b,a\oplus c,b),\quad
d\in\{1,2,\ldots,63\},
$$
then \eqref{Eq.BASH-S3.1} becomes
$$
w_0 \wedge (w_0\oplus w_0^d \oplus w_1) \oplus
w_1 \wedge (w_0\oplus w_2) \oplus
w_2 \wedge (w_1) =
w_0 \wedge (w_0\oplus w_0^d)\neq 11\ldots 1.
$$

Thus, we found the following solution for the problem:
$$
S3(a,b,c)\oplus L(a,b,c)=
(b \vee \neg c\oplus a^d\oplus b,
a \vee c\oplus a \oplus b\oplus c,
a \wedge b\oplus b\oplus c).
$$

Note that there are many other possible solutions.

This problem was completely solved by three participants in the first round and by nine teams in the second round. Many of these solutions were interesting and compact.

\subsection{Problem ``Metrical cryptosystem -- 2''}

\subsubsection{Formulation}

Let $\mathbb{F}_2^n$ be an $n$-dimensional vector space over the field $\mathbb{F}_2=\{0,1\}$.
Alice and Bob exchange messages using the following cryptosystem.
\begin{itemize}[noitemsep]
\item[{\bf 1.}]
 First, they use a supercomputer to calculate two special large secret sets $A,B\subseteq\mathbb{F}_2^n$ which have the following property: there exists a constant $\ell$ ($\ell\geqslant 26$), such that for any $x\in\mathbb{F}_2^n$ it holds $$d(x,A)+d(x,B)=\ell,$$ where $d(x,A)$ denotes Hamming distance from the vector $x$ to the set $A$.

\item[{\bf 2.}]
 Alice then saves the number $\ell$, the set $A$ and a set of vectors $a_1,a_2,\ldots,a_r$ such that for any $k: 0\leqslant k \leqslant \ell$, there is a vector $a_i$ at distance $k$ from $A$. Similarly, Bob saves the number $\ell$, the set $B$ and a set of vectors $b_1,b_2,\ldots,b_s$ such that for any $k: 0\leqslant k \leqslant \ell$, there is a vector $b_i$ at distance $k$ from $B$.

\item[{\bf 3.}]
  Text messages are encrypted letter by letter. In order to encrypt a letter Alice replaces it  with its number in the alphabet, say $k$. Then she chooses some vector $a_i$ at distance $k$ from the set $A$ and sends this vector over to Bob. Bob then calculates the distance $d(a_i,B)$ and using the property of the sets $A,B$, calculates $k = \ell - d(a_i,B)$. So, he gets the letter Alice sent.
  If Bob wants to send an encrypted message to Alice, he does the same but using his saved vectors and the~set~$B$.

\end{itemize}

Eve was able to hack the supercomputer when it was calculating the sets $A$ and~$B$. She extracted the set $C$ from its memory, which consists of all vectors of $\mathbb{F}_2^n$ that are at distance 1 or less from either $A$ or $B$. She also learned that $\ell$ is even.

Help Eve to crack the presented cryptosystem (to decrypt any short intercepted message)! You know that she has an (illegal) access to the supercomputer, which can calculate and output the list of distances from all vectors of $\mathbb{F}_2^n$ to any input set $D$ in reasonable (but not negligible) time.

\textbf{Remarks.} Recall several definitions and notions. The {\it Hamming distance} $d(x,y)$ between vectors $x$ and $y$ is the number of coordinates in which these vectors differ. Distance from vector $y\in\mathbb{F}_2^n$ to the set $X\subseteq\mathbb{F}_2^n$ is defined as $d(y,X)=\min_{x\in X} d(y,x)$.

\subsubsection{Solution}

Let us denote by $A_i$ ($B_i$ respectively) the set of all vectors at distance $i$ from the set $A$ ($B$ respectively):
$$A_i = \{x\in \mathbb{F}_2^n : d(x,A) = i\}, \ \  B_i = \{x\in \mathbb{F}_2^n : d(x,B) = i\}.$$
It is easy to see that
\begin{itemize}[noitemsep]
    \item{$A=A_0=B_{\ell}$,}
    \item{$B=B_0=A_{\ell}$,}
    \item{$A_i=B_{\ell - i}$ for any $i\in \{0,\ldots,\ell\}$,}
    \item{$C = A_1 \cup B_1 \cup A_0 \cup B_0$.}
\end{itemize}

From the definition of the Hamming distance it is easy to prove that if a vector $x$ lies in the set $A_i$, then it is at distance $|i-j|$ from the set $A_j$ for any $i,j$.

\begin{proof}
Indeed, if $i=j$, the statement is trivial.

Assume that $i>j$. By definition, $d(x,A)=i$, so there exists a shortest path of length $i$ from $A$ to $x$, consisting of vectors $x_0,x_1,\ldots,x_i=x$, where $x_0\in A$. Since consecutive vectors in the path differ in only one coordinate, and vectors from $A_s$ and $A_t$ can be neighbours only if $|s-t| \leqslant 1$, it follows that $x_k\in A_k$ for every $k=0,\ldots,i$. So, vector $x_j$ from the path belongs to $A_j$ and is at distance $i-j$ from vector $x$. Therefore, $d(x,A_j)\leqslant i-j$. Distance cannot be less than $i-j$, because then $d(x,A)$ would have been less that $i$, which contradicts conditions of the statement. Thus, $d(x,A_j) = i-j$.

If $i<j$, then we can replace $A_i$ with $B_{\ell-i}$, $A_j$ with $B_{\ell-j}$ and use $B$ instead of $A$ for the same argument as in the previous case.
\end{proof}

In particular, given $x$ is in $A_i$, it is at distance $|i-1|$ from the set $A_1$ and at distance $|i-(\ell-1)|$ from the set $B_1$.

Let us ``feed'' the set $C$ to the supercomputer. We denote the maximal distance from vectors of $\mathbb{F}_2^n$ to vectors of $C$ as $r$, and the set of all vectors achieving this distance as $\widehat{C}$. Taking into account the statement proven above (and the fact that $\ell$ is even), we can see that the maximum is achieved for vectors of the set $A_{\frac{\ell}{2}}$. Hence, $r = \frac{\ell}{2} - 1$ and $\widehat{C} = A_{\frac{\ell}{2}}$. Thus, we can calculate $\ell$ as $2r+2$.

Assume now that Alice sends a message $a_{i_1},a_{i_2},\ldots,a_{i_k}$ to Bob. Eve intercepts it and (using the obtained table of distances from the set $C$) calculates that these vectors are at distances $s_1,s_2,\ldots,s_k$ from the set $C$. Therefore, they are at distances $s_1+1,s_2+1,\ldots,s_k+1$ from the set $A\cup B$. Since $d(x, A\cup B) = \min(d(x, A),d(x, B))$, each encrypted letter could be either $s_i+1$ or $\ell-(s_i+1)$. If one of these two numbers is greater than 26, we can easily determine the encrypted letter, if not, we can consider both possibilities.
In the worst case we would need to consider $2^N$ variants, where $N$ is the length of the message, but since messages are short and are written in natural language, we do not need to check all of them and the decryption should not be hard.

\textbf{Note:} Sets $A$ and $B$ satisfying condition from Step 1 of the problem (for an arbitrary constant~$\ell$ not necessarily greater than $26$) are called \emph{strongly metrically regular} and are studied in \cite{Obl18}.

Best solutions to the problem were submitted by Alexey Chilikov (Bauman Moscow State Technical University, Russia) and Saveliy Skresanov (Novosibirsk State University, Russia).

\subsection{Problem ``A fixed element''}

\subsubsection{Formulation}

A polynomial $f(X_1,\dots,X_n)\in\Bbb F_2[X_1,\dots,X_n]$ is called {\em reduced} if the degree of each $X_i$ in $f$ is at most $1$. For $0\leqslant r\leqslant n$, the $r$th order Reed~---~Muller code of length~$2^n$, denoted by $R(r,n)$, is the $\Bbb F_2$-space of all reduced polynomials in $X_1,\dots,X_n$ of total degree less or equal than $r$. We also define $R(-1,n)=\{0\}$.

The general linear group $\text{GL}(n,\Bbb F_2)$ acts on $R(r,n)$ naturally:
Given $A\in\text{GL}(n,\Bbb F_2)$ and $f(X_1,\dots,X_n)\in R(r,n)$, $Af$ is defined to be the reduced polynomial obtained from $f((X_1,\dots,X_n)A)$ by replacing each power $X_i^k$ ($k\geqslant 2$) with~$X_i$. Consequently, $\text{GL}(n,\Bbb F_2)$ acts on the quotient space $R(r, n)/R(r-1,n)$.

Let $A\in\text{GL}(n,\Bbb F_2)$ be such that its characteristic polynomial is a primitive irreducible polynomial over $\Bbb F_2$. Prove that the only element in $R(r, n)/R(r-1,n)$, where $0 < r < n$, fixed by the action of $A$ is~$0$.

\subsubsection{Solution}

Let $\binom{\{1,\dots,n\}}r$ denote the set of $r$-subsets of $\{1,\dots,n\}$. When $A$ acts on $R(r, n)/R(r-1,n)$, its matrix with respect to the basis $\prod_{i\in I}X_i$, $I\in\binom{\{1,\dots,n\}}r$, is the $r$th compound matrix $C_r(A)$ of $A$. The eigenvalues of $A$ are $\gamma^{2^i}$, $0\le i\le n-1$, where $\gamma$ is a primitive element of $\Bbb F_{2^n}$. The eigenvalues of $C_r(A)$ are all possible products of $r$ eigenvalues of $A$, i.e.,
\[
\gamma^{\sum_{i\in I}2^i},\quad I\in \binom{\{0,\dots,n-1\}}r.
\]
Clearly, the above expression never equals $1$. Hence $1$ is not an eigenvalue of $C_r(A)$. Therefore, the action of $A$ does not fix any nonzero element in $R(r, n)/(r-1,n)$.

The problem was solved by four teams in the second round: Aleksei Udovenko (University of Luxembourg), the team of Dianthe Bose and Neha Rino (Chennai Mathematical Institute, India), the team of Andrey Kalachev, Danil Cherepanov and Alexey Radaev	(Bauman Moscow State Technical University, Russia), the team of Sergey Titov and Kristina Geut	(Ural State University of Railway Transport, Russia).

\subsection{Problem ``Hash function \texttt{FNV-1a}''}

\subsubsection{Formulation}

Hash function
\href{http://www.isthe.com/chongo/tech/comp/fnv/}{\textcolor{blue}{{\tt FNV-1a}}} \cite{FNV}
processes a message~$x$ composed of bytes
$x_1,x_2,\ldots,x_n\in\{0,1,\ldots,255\}$ in the following way:
\begin{enumerate}[noitemsep]
\item[{\bf 1.}]
$h\leftarrow h_0$;
\item[{\bf 2.}]
for $i=1,2,\ldots,n$:
$h\leftarrow (h\oplus x_i)g\bmod 2^{128}$;
\item[{\bf 3.}]
return~$h$.
\end{enumerate}
Here~$h_0=144066263297769815596495629667062367629$,
$g=2^{88}+315$. The expression $h\oplus x_i$ means that the least significant
byte of $h$ is added bitwise modulo~$2$ with the byte~$x_i$.

Find a collision, that is, two different messages~$x$ and~$x'$ such
that~$\text{{\tt FNV-1a}}(x)=\text{{\tt FNV-1a}}(x')$.
Collisions on short messages and collisions that are obtained without
intensive calculations are welcomed. Supply your answer as a pair of two
hexadecimal strings which encode bytes of colliding messages.

\subsubsection{Solution}

We will base on the solution for the problem ``{\tt FNV2}'' (NSUCRYPTO'2017) \cite{nsucrypto-2017}, where it was required to find a collision for the similar hash function {\tt FNV2}.
{\tt FNV-1a} differs from {\tt FNV2} in the following: instead of the $\oplus$ operation for adding $h$ and $x_i$ it uses standard $+$ operation.

It is easy to see that
$$
\text{{\tt FNV2}}(x_1 x_2\ldots x_n)=
(h_0 g^n + x_1 g^n+x_2 g^{n-1}+\ldots + x_n g)\bmod 2^{128}.
$$

For  {\tt FNV2}, we found a relation
$$
a_1 g^{n-1}+ a_2 g^{n-2} + \ldots + a_n \equiv 0\!\!\pmod{2^{128}},
$$
where~$a_i\in\{-255,\ldots,255\}$.

Then we represented $a_i$ as the difference $x_i-x_i'$ and found a collision
$$
\text{{\tt FNV2}}(x_1 x_2\ldots x_n)-\text{{\tt FNV2}}(x_1' x_2'\ldots x_n')=
a_1 g^n +a_2 g^{n-1}+\ldots + a_n g\equiv 0\!\!\pmod{2^{128}}.
$$

Let us call a representation $a_i=x_i-x_i'$ as a {\it splitting} of $a_i$. There can be several splittings for a given $a_i$. Each of them induces two trajectories of intermediate values of $h$: the trajectory starting with a message $x_1x_2\ldots x_n$ and
the trajectory starting with a message~$x_1'x_2'\ldots x_n'$.

Let~$h_i$ and~$h_i'$~be the low bytes of~$h$ for the first and second trajectories respectively before the additions $h+x_i$ and~$h+x_i'$.
Let us call a splitting {\it suitable} if
$$
h_i+x_i< 256,\quad
h_i'+x_i'< 256,\quad
i=1,2,\ldots,n.
$$

Let us evaluate the probability of existing a suitable splitting for $a_i$.
We will assume that $h_i$, $h_i'$ are realizations of independent random variables with uniform distribution over $\{0,1,\ldots,255\}$.

Bytes $x_i$ and $x_i'$ can take any value from intervals $\{0,\ldots,255-h_i\}$ and $\{0,\ldots,255-h_i'\}$ respectively.
At the same time, the difference $x_i-x_i'$ takes value from the interval
$\{-255+h_i',\ldots,255-h_i\}$.

Then $a_i$ is in the interval $\{-255+h_i',\ldots,255-h_i\}$ with the probability
$$
\Pr{(-255+h_i'\leqslant a_i\leqslant 255-h_i)}=
\begin{cases}
\Pr{(h_i\leqslant 255-a_i)}, & a_i\geqslant 0,\\
\Pr{(h_i'\leqslant 255-|a_i|)}, & a_i< 0,
\end{cases}
$$
that is equal to $1-|a_i|/256$.

Thus, the probability of existing a suitable splitting for the whole sequence $a_1a_2\ldots a_n$ is the following:
$$
p=\prod_{i=1}^n
\left(1-\frac{|a_i|}{256}\right).
$$
This probability can be rather high. For example, $p\approx 1/25$
for the following sequence for~$n=18$:
$$
(-64,5,73,35,-53,19,-10,-78,-44,48,61,-1,-80,26,-22,72,-31,0).
$$
Or,  $p\approx 1/13$ for the following sequence for $n=19$:
$$
(-37,34,-74,-4,-17,33,-18,21,54,33,-1,58,-71,-13,-10,11,-88,-19,0).
$$

Moreover, the probability can be increased if we change a strategy of finding suitable splittings. We can allow to modify splittings $a_1,\ldots,a_{i-1}$ that have been already built if it is impossible to find a splitting for $a_i$.

After finding a suitable splitting, we determine the sequences $(h_i)$, $(h_i')$. Then we determine the bytes $\tilde{x}_i$, $\tilde{x}_i'$ such that
$$
h_i\oplus \tilde{x}_i=h_i+x_i,\quad
h_i'\oplus \tilde{x}_i'=h_i'+x_i',\quad
i=1,2,\ldots,n.
$$

It is important that there are no carries in high bytes in additions $h_i+x_i$, $h_i'+x_i'$; and $\tilde{x}_i$, $\tilde{x}_i'$ can be always found.
Then a collision for {\tt FNV-1a} is a pair of messages
$\tilde{x}_1\tilde{x}_2\ldots \tilde{x}_n$ and
$\tilde{x}_1'\tilde{x}_2'\ldots \tilde{x}_n'$.

It remains to say that the sequence~$(a_i)$ can be found using LLL algorithm. The algorithm is applied to the lattice defined by the basic vectors
\begin{align*}
\mathbf{b}_1 &= (1,0,\ldots,0,g^{n-1}\bmod 2^{128}),\\
\mathbf{b}_2 &= (0,1,\ldots,0,g^{n-2}\bmod 2^{128}),\\
&\ldots\\
\mathbf{b}_n &= (0,0,\ldots,1,g^0\bmod 2^{128}),\\
\mathbf{b}_{n+1} &= (0,0,\ldots,0, t 2^{128}),
\end{align*}
where~$t$~is a small integer.
LLL finds a short basis of the lattice, i.\,e. vectors
$$
v=\sum_{i=1}^{n+1}a_i \mathbf{b}_i
$$
with small coordinate values. Let the last coordinate~$v$ equal to $0$. Then
$$
\sum_{i=1}^{n+1}a_i g^{n-i}\equiv 0\!\!\pmod{2^{128}},
$$
i.\,e. $(a_1,\ldots,a_n)$ is a required solution.

This problem was completely solved by fourteen teams (the most of them used a reduction to the problem {\tt FNV2}).
Some examples of collisions proposed by participants (in HEX format) are given in Table~\ref{tab-hash}.

\begin{table}
\caption{Collisions of {\tt FNV-1a}.}
\label{tab-hash}
\smallskip
\begin{tabular}{|c|c|}
\cline{1-2}
{\tt message 1} & {\tt message 2}\\
\cline{1-2}
{\tt f1dd5921afd29cbd33b357184e8c} & {\tt 928ea41b7373792aae2bfa72ca64}\\
\cline{1-2}
{\tt eb18151b160aa95e0511357e158b58} & {\tt ab775b3a7c7c7c7c7c7c3a5dc94e}\\
\cline{1-2}
{\tt f828e4070672220b195e0ddd2114a4c008} & {\tt 3638fa655d1b61e21419134803222bbb35}\\
\cline{1-2}
{\tt 3a7a3a7a3a4a5a5a5a5a5a5a5a5a5a5a5a5a} & {\tt 51089c5e7fe7cc2d740b5f70b3cb5461824d}\\
\cline{1-2}
{\tt f4331cede51639057d05f80f1d6638b40b286f} & {\tt eb270505187332116c611402081f1155326013}\\
\cline{1-2}
{\tt 07160c2e0b700b1338ef6e63360419060507} & {\tt 10610bf23b0573e2317106176a171c6a4c6e}\\
\cline{1-2}
{\tt 00ca0000cb000000000000000029092100d814} & {\tt 2d000158001b773a6364fc0905000000e90000}\\
\cline{1-2}
\end{tabular}
\end{table}

\subsection{Problem ``\texttt{TwinPeaks2}''}

\subsubsection{Formulation}

Bob realized that his cipher from last year, \texttt{TwinPeaks} (NSUCRYPTO'2017) \cite{nsucrypto-2017}, is not secure enough and modified it. He considerably increased the number of rounds and made rounds more complicated. Bob's new cipher works as follows.

A message $X$ is represented as a binary word of length $128$. It is
divided into four 32-bit words $a,b,c,d$ and then the following round
transformation is applied 48 times:
$$
(a,b,c,d)\leftarrow
(b, c, d, a \oplus S_3(S_1(b)\oplus S_2(b\wedge \neg c \oplus c \vee d)\oplus S_1(d))),
$$
Here $S_1,S_2,S_3$ are secret permutations over $32$-bit words;
$\neg$, $\wedge$, $\vee$, $\oplus$ are binary bitwise ``NOT', ``OR'', ``AND'',
``XOR'' respectively (the operations are listed in descending order of
priority). The concatenation of the final $a,b,c,d$ is the resulting ciphertext
$Y$ for the message $X$.

Agent Cooper again wants to read Bob's messages!
He intercepted the ciphertext
$$Y=\texttt{DEB239852F1B47B005FB390120314478}$$
and  also captured Bob's
smartphone with the \texttt{TwinPeaks2} implementation!
\href{https://nsucrypto.nsu.ru/olymp/2018/round/2/task/4}{\textcolor{blue}{Here}}
it is \cite{twinpeaks2}. Now Cooper (and you too) can encrypt any messages with
\texttt{TwinPeaks2} but still can not decrypt any.
Help Cooper to decrypt $Y$.

{\bf Remarks.}
The ciphertext is given in hexadecimal notation, the first byte is \texttt{DE}.

\subsubsection{Solution}

Let $F$ be the round transformation of \texttt{TwinPeaks2}:
$$
F(a,b,c,d)=
(b, c, d, a \oplus f(b,c,d)).
$$
The encryption transformation is the composition of 48 copies of $F$,
i.\,e. it can be written as $F^{48}$. Consequently, $F^{-48}$ is the decryption transformation.
Let
$$
\tau(a,b,c,d)=(d,c,b,a).
$$
Let us note that $f(b,c,d) = f(d,c,b)$. Then the composition of $F$, $\tau$ and $F$ gives us $\tau$:
$$
F\circ\tau\circ F(a,b,c,d)=
F(a \oplus f(b,c,d),d,c,b)=
F(a \oplus f(d,c,b),d,c,b)=
(d,c,b,a).
$$
Hence,
$$
F^{48}\tau F^{48}=\tau
$$
or
$$
F^{-48}=\tau F^{48}\tau^{-1}=\tau F^{48}\tau.
$$

Thus, in order to decrypt $Y$ one should write its 32-bit blocks in reverse order, encrypt the result and then reverse the order of the blocks again.
The result will be a hexadecimal word, which gives us the desired message
$$
\texttt{attacksgetbetter}.
$$

The best solution to the problem has been submitted by Carl L{\"o}ndahl (Sweden), which not only provides a clean theoretical solution, but also proposes a slide attack on the cipher.

\subsection{Problem ``An Enigmatic Challenge''}

\subsubsection{Formulation}

\begin{figure}[h]
\centering
\includegraphics[width=0.8\textwidth]{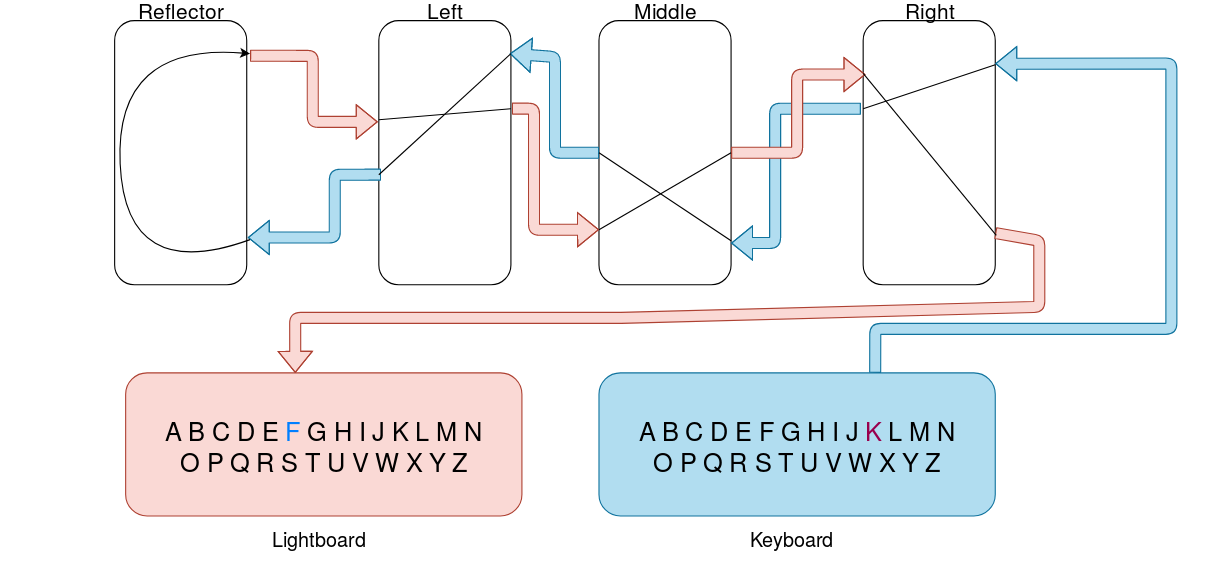}
\vspace{-0.3cm}
\caption{{\small A simplified Enigma}}
\label{fig:Enigma}
\end{figure}

The Enigma machine is a symmetric cipher famous for being used during the
Second World War by the German military.
Its internal structure comprises a 26-letter Latin alphabetic permutation,
implemented as rotors.
The machine used for this problem consists of 3 \textbf{rotors} and a
\textbf{reflector}.

Figure~\ref{fig:Enigma} shows how a simplified Enigma machine works.
The key components are the set of input switches (2) -- which are reduced to 4 in the
example but could have been 26 for the Latin alphabet --
an input plugboard (3,7,8), three rotors (5), the reflector (6)
and the output board (9).

The components have the following functionality:

    $\bullet$ \textbf{Rotors:} a rotor (5) is a wheel with the upper-case alphabet in order on the rim and a
    hole for an axle. On both sides of a rotor are 26 electrical contacts each under a letter.
    Each contact on one side is wired to a contact on the other side at a different position.
    The rotor implements an one-to-one and onto function between the upper-case letters,
    where each letter is mapped to a different one (an irreflexive permutation).

 $\bullet$   \textbf{Reflector:} the reflector (6) is positioned after the rotors and has contacts
    for each letter of the alphabet on one side only.
    The letters are wired up in pairs, so that an input current
    on a letter is reflected back to a different letter.

\textbf{The input message:}
is permuted by the rotors, passes through the reflector and then goes back
through the rotors in the reverse order (as depicted in the figure).
Finally, the light bulb indicates
the encrypted letter. The plugboard plays no role in permuting the letter
for this challenge, although it could have.

To prevent simple frequency analysis attack the Right rotor rotates with every
new input. After the Right rotor completed a full rotation (after 26 letters were encrypted),
the Middle rotor rotates once. Similarly, after the Middle rotor completes a full rotation
(and the Right rotor complete 676 rotations), the left rotor rotates once\footnote{This
means that an input letter is processed, in order, by three permutation --
Right, Middle and Left -- reflected by the reflector, and processed once again,
in order, by the inverse permutations corresponding to Left, Middle and Right rotors
before being output. Once the letter passes through a rotor, it is permuted with one position, the rotor's permutation is applied, and the result goes directly into the following rotor, which acts similarly.}.

\textbf{Challenge:} you will play the role of an attacker that knows the source of the
plaintext to be encrypted. You are given a ciphertext corresponding to a plaintext
taken from this known source which happens to be
``Moby Dick'' by Herman Melville, and you are asked to recover the plaintext.
The plaintext consists only of trimmed capital letters with no punctuation marks and spaces
and is contiguous. All letters are from the Latin alphabet.
Extra information on the settings of the rotors is provided:
the configuration of the first rotor is very close to the one used in the 1930 commercial
version (that was {\tt EKMFLGDQVZNTOWYHXUSPAIBRCJ}).

\textbf{Ciphertext:}
\begin{center}
\noindent
{\tt
 RHSM ZHXX AOWW ZTWQ QQMB CRZA BARN MLAV MLSX SPBA ZTHG \\YLGE VGZG KULJ FLOZ RQAW YGAA DCJB YWBW IYQQ FAAO RAGK \\BGSW OARG EYSP IKYE LLUO YCNH HDBV AFKD HETA ONNR HXHE \\BBRT ROZD XJCC OMXR PNSW UAZB TNJY BANH FGCS GJWY YTBV\\ VGLX KUZW PARO NMXP LDLZ ICBK XVSJ NXCF SOTA AQYS YZFX\\ MZDH MSZI ABAH RFXT FTPU VWMC PEXQ NZVA LMFX BHKG QGYS \\BIYE MEUE PJNR AVTL JSUZ PLHQ MOUI IQFD HVXI NOOJ YJAF\\ WAVU PVQA FMKP AHLK XJYD GITB QSPK CUZU XPRK MUJJ YRJ~
}
\end{center}

Link to \href{https://gist.githubusercontent.com/StevenClontz/4445774/raw/1722a289b665d940495645a5eaaad4da8e3ad4c7/mobydick.txt}{\textcolor{blue}{``Moby Dick''}} text file can be found in \cite{moby-dick}.

\subsubsection{Solution}

It is easy to observe that the Left and Middle Rotors will not change for each block of 26 characters of the plaintext. From this point of view, we can regard the composition of permutations induced (in order) by the Middle and Left rotors, the reflector and as well as the inverses of the Left and Middle rotors, as one, fixed permutation. After the next 26 letters are processed, the Middle rotor turns, and a distinct permutation is to be used for the incoming block of 26 letters. Due to the fact the challenge ciphertext is less than 676 characters, we do not bother with turning the Left rotor.

To fix some notations, let $\pi_i, L_i, M_i$ denote permutations defined on the
 set $\{{\tt A}, \ldots, {\tt Z}\}$.
%----Added----%
If $L:\{{\tt A}, \ldots, {\tt Z}\} \to \{{\tt A}, \ldots, {\tt Z}\}$ denotes the
permutation defining the Left rotor, by
$L_i:\{{\tt A}, \ldots, {\tt Z}\} \to \{{\tt A}, \ldots, {\tt Z}\}$ and
$L_i = L \circ Rot_i$, we represent the action of applying the Left rotor over
the alphabet, where $Rot_i$ represents the alphabet's rotation by $i$.
We use a similar notation for $M_i$,
with $i$ denoting each block of 26 letters to be processed.
%----Added----%
That is $i \in \{1, \ldots, \lceil{\frac{|C|}{26}}\rceil \}$, where $|C|$ denotes the length of the ciphertext (its number of characters). We also write
$$
    \pi_i =  M_i^{-1} \circ L_0^{-1} \circ \rho \circ L_0 \circ M_i,~~
                i \in \Big\{1, \ldots, \Big\lceil{\frac{|C|}{26}}\Big\rceil \Big\}.
$$

The next step is to split the challenge ciphertext into blocks of 26 characters,
and use the fact that for each block $i$, $\pi_i$ acts as an oracle that returns
the same value for the same input. We will correlate this with the information
that is \textit{a priori} given on the first rotor. Although we do not have its
exact configuration, we use the fact that the unknown rotor is close
to a known one ({\tt EKMFLGDQVZNTOWYHXUSPAIBRCJ} --- commercial Enigma 1930).
The configuration used for this problem permutes 4 elements amongst
the ones of the 1930 configuration and
then applies a circular permutation of length four.

The permutation corresponding to the given Right rotor of the commercial
Enigma (1930) is the following:
\begin{center}
{\tt
    A~B~C~D~E~F~G~H~I~J~K~L~M~N~O~P~Q~R~S~T~U~V~W~X~Y~Z

    E~K~M~F~L~G~D~Q~V~Z~N~T~O~W~Y~H~X~U~S~P~A~I~B~R~C~J
}
\end{center}

Then we take the first block of 26 letters and obtain their inverses, minding
the fact that the rotor shifts with one place to the left after we read one
letter. Hence, for the first block  we obtain:
\begin{center}
{\tt
RHSM~ZHXX~AOWW~ZTWQ~QQMB~CRZA~BA

UVAH~\textcolor{red}{F}O\textcolor{red}{F}\textcolor{blue}{L}~VRDQ~TDNG~DQ\textcolor{blue}{L}A~BOIR~JJ
}
\end{center}

Now we remark on a ``distance-preserving" property: if the distance between identical characters returned by $\pi_i$ (the input to the Right rotors) is $\ell$, then it maintained in the original plaintext. As an example, the group {\tt ZHXX} in the first block of ciphertext has been obtained for the group {\tt FOFL} and we note a distance of 2 ({\tt F} $\to$ {\tt O} $\to$ {\tt F}) between {\tt F} and {\tt F}. This means that an alphabetical distance of $2$ exists between the corresponding letters of the plaintext. More precisely,~if:
$$
    \pi_1\big( R(x) \big)  =  \pi_1\big( R'(y) \big)~,
$$
where $R'$ is obtained by shifting $R$ with $\ell$ elements, then the character $y$ is at a distance of $\ell$ from the character $x$ (but in the opposite sense). Based on this observation, the solution is to identify such pairs inside a block and record the distance between them. As $4$ elements are permuted in the real configuration of the rotor, false positives will appear.

After the colliding characters per block, say in position $i$ and $j$, have been identified and their distance recorded, say the distance is $\ell$, one will simply write a script that will pass through the given plaintext (after removing the non-alphabetic characters) and identify the sequence (matching the length of the ciphertext) where the distance between the characters in position $i$ and $j$ is $\ell$.

Finally, the plaintext that is to be recovered is:

\begin{displayquote}
\textit{
  ALREADY we are boldly launched upon the deep ; but\newline
        soon we shall be lost in its unshored, harbourless immen-\newline
        sities. Ere that come to pass ; ere the Pequod'a weedy\newline
        hull rolls side by side with the barnacled hulls of the\newline
        leviathan; at the outset it is but well to attend to a\newline
        matter almost indispensable to a thorough appreciative\newline
        understanding of the more special leviathanic revelations\newline
        and allusions of all sorts which are to follow.
}
\end{displayquote}

Finally, eight teams completely solved the problem. Note, that many teams used a simple method that almost completely determined the plaintext. It is based on the fact that no letter from the plaintext gets mapped to the same letter in the ciphertext using Enigma. But this approach gives two possible solutions and does not allow one to prove that one of them is not correct.

\subsection{Problem ``Orthogonal arrays'' (unsolved)}

\subsubsection{Formulation}

Orthogonal arrays are closely connected with cryptographic Boolean functions. Namely, supports of correlation immune functions give orthogonal arrays when their elements are written as the rows of an array.

Given three positive integers $n$, $t$ and $\lambda$ such that $t<n$, we call a $\lambda 2^t\times n$ binary array (i.e. matrix over the 2-element field) a $t-(2,n,\lambda)$ {\it orthogonal array} if in every subset of $t$ columns of the array, every (binary) $t$-tuple appears in exactly $\lambda$ rows.
$t$ is called the strength of this orthogonal array.

Find a $4-(2,11,\lambda)$ orthogonal array with minimal value of $\lambda$.

\subsubsection{Solution}

The best known answer to this question is $\lambda = 8$ \cite{OA-2}, but it is unknown whether there exists a $4-(2,11,\lambda)$ orthogonal array for $\lambda < 8$. This open problem remains unsolved.
Participants suggested several ideas.

The most interesting one was proposed by Aleksei Udovenko (University of Luxembourg).
His study starts with the Nordstrom --- Robinson code (that is, the Kerdock code of length 16 and size 256, whose dual distance is the minimum distance of the Preparata code, that is 6, which gives a strength of the orthogonal array (OA) equal to 5). Only the codewords with
the first element equal to zero are kept and their coordinate at 0 is deleted, which makes size 128, length 15 and strength 4. Then three columns are erased from the OA, which does not reduce the strength, and the resulting OA provides a solution to the problem with $\lambda=8$. It is then shown (by using known results) that, for any solution to the problem, $\lambda$ is at least 6. This is interesting. The solution found is written in the form $(x,F(x))$ where $F$ is a quadratic $(7,4)$-function. Its determination allows one to determine the 4-th order correlation immune function whose support is this OA. This is a 11-variable Boolean function of algebraic degree 5. Then the annihilators of this function are studied. It is shown that the function has a linear annihilator (and has then algebraic immunity 1). After an observation on the impossibility of extending a solution which would have $\lambda\leqslant 7$, the Xiao --- Massey characterization of OA is proved again in different terms. It is also shown that any affine annihilator of a $t$-th order correlation immune function must be $t$-resilient which is a nice observation. A computer search is made with Integer Linear Programming showing that any 4-th order correlation immune function having an affine annihilator should have weight at least 128, which is a nice observation. This nice work concludes with open questions.

Another good solution was given by the team of Evgeniya Ishchukova, Vyacheslav Salmanov and Oksana Shamilyan (Southern Federal University, Russia). They first studied the maximum value of $n$, given $t$, for small values of $t$. Then an algorithm was designed which reduces the search to solutions having some symmetries observed in smaller values of $t$ and $n$. Finally, a solution was given with $\lambda=8$ which is the coset of a linear code of length $n=11$ and dimension $k=7$ (and therefore 128 codewords), with dual distance 5 and the corresponding function is then indeed 4th order correlation immune giving a $4-(2,11,\lambda)$ orthogonal array. Unfortunately, the question whether 128 is minimal was not addressed.

\subsection{Problem ``Sylvester matrices'' (unsolved)}

\subsubsection{Formulation}

Sylvester matrices play a role in security since they are connected with topics like secret sharing and MDS codes constructed with cellular automata.

Consider two univariate polynomials over the 2-element field, $P_1(x)$ of degree $m$ and $P_2(x)$ of degree $n$, where $P_1(x) = a_mx^m + \ldots + a_0$ and $P_2(x) = b_nx^n + \ldots + b_0$. The {\it Sylvester matrix} is an $(m+n)\times(m+n)$ matrix formed by filling the matrix beginning with the upper left corner with the coefficients of $P_1(x)$, then shifting down one row and one column to the right and filling in the coefficients starting there until they hit the right side. The process is then repeated for the coefficients of $P_2(x)$. All the other positions are filled with zero.

Let $n>0$, $m>0$. Prove whether there exist $(m+n)\times(m+n)$ invertible Sylvester matrices whose inverses are Sylvester matrices as well.

{\bf Example.}
For $m=4$ and $n=3$, the Sylvester matrix is the following:
$$
\left(
  \begin{array}{ccccccc}
    a_4 & a_3 & a_2 & a_1 & a_0 & 0 & 0 \\
    0 & a_4 & a_3 & a_2 & a_1 & a_0 & 0 \\
    0 & 0 & a_4 & a_3 & a_2 & a_1 & a_0 \\
    b_3 & b_2 & b_1 & b_0 & 0 & 0 & 0 \\
    0 & b_3 & b_2 & b_1 & b_0 & 0 & 0 \\
    0 & 0 & b_3 & b_2 & b_1 & b_0 & 0 \\
    0 & 0 & 0 & b_3 & b_2 & b_1 & b_0 \\
  \end{array}
\right)
$$

\subsubsection{Solution}

We are pleased to say that three teams completely solved this problem! They are Alexey Chilikov (Bauman Moscow State Technical University, Russia), the team of Radu Caragea, Madalina Bolboceanu and Miruna Rosca	 (Bitdefender, Romania), the team of Samuel Tang and Harry Lee (Hong Kong). Here we present the main idea for the solution.

\underline{Case 1: $m\leqslant n$}. Let $P_1(x)=x^m$ and $P_2(x)=x^n+1$. Then their Sylvester matrix is the following:
\begin{equation*}
\left(
\begin{array}{ccccccc}
\multicolumn{2}{c}{{\bf I}_n} & \multicolumn{1}{|c}{{\bf 0}_{n\times m}} \\ \cline{1-3}
\multicolumn{1}{c|}{{\bf I}_m} & \multicolumn{1}{c|}{{\bf 0}_{m\times(n-m)}} &{\bf I}_m
\end{array}
\right),
\end{equation*}
where ${\bf I}_k$ denotes the $k\times k$ identity matrix; and ${\bf 0}_{k\times \ell}$ is the $k\times \ell$ zero matrix. Taking all operations over the the 2-element field, it is clear that
\begin{equation*}
\left(
\begin{array}{ccccccc}
\multicolumn{2}{c}{{\bf I}_n} & \multicolumn{1}{|c}{{\bf 0}_{n\times m}} \\ \cline{1-3}
\multicolumn{1}{c|}{{\bf I}_m} & \multicolumn{1}{c|}{{\bf 0}_{m\times(n-m)}} &{\bf I}_m
\end{array}
\right)\cdot\left(
\begin{array}{ccccccc}
\multicolumn{2}{c}{{\bf I}_n} & \multicolumn{1}{|c}{{\bf 0}_{n\times m}} \\ \cline{1-3}
\multicolumn{1}{c|}{{\bf I}_m} & \multicolumn{1}{c|}{{\bf 0}_{m\times(n-m)}} &{\bf I}_m
\end{array}
\right)=\left(
\begin{array}{ccccccc}
\multicolumn{1}{c|}{{\bf I}_n} & {\bf 0}_{n\times m} \\ \cline{1-2}
\multicolumn{1}{c|}{{\bf 0}_{m\times n}} & {\bf I}_{m}
\end{array}
\right)={\bf I}_{m+n}.
\end{equation*}
Thus, the considered Sylvester matrix is an involutory matrix. Therefore, its inverse is the Sylvester matrix as well.

\underline{Case 2: $m>n$}. Assume that the inverse of the Sylvester matrix of $P_1(x)$ and $P_2(x)$ is also the Sylvester matrix for two polynomials over the 2-element field, say $Q_1(x)=c_px^p+c_{p-1}x^{p-1}+\dots+c_0$, $Q_2(x)=d_qx^q+d_{q-1}x^{q-1}+\dots+d_0$, of degrees $p>0$ and $q>0$ respectively, which satisfy $p+q=m+n$. The product of Sylvester matrices which correspond to $P_1(x),P_2(x)$ and $Q_1(x),Q_2(x)$ is equal to ${\bf I}_{m+n}$, in particular
\begin{equation*}
\left(
\begin{array}{cccccccc}
a_m & a_{m-1} & \dots & a_0 \\
 & a_m & a_{m-1} & \dots & a_0 \\
 & & & \ddots & \ddots \\
 & & & a_m & a_{m-1} & \dots & a_0 \\ \cline{1-7}
b_n & b_{n-1} & \dots & b_0 & 0 & \dots & 0 \\ \cline{1-7}
& b_n & b_{n-1} & \dots & b_0 \\
 & & & \ddots & \ddots \\
 &  &  & b_n & b_{n-1} & \dots & b_0
\end{array}
\right)\cdot\left(
\begin{array}{cc}
c_p \\
0 \\
\vdots \\
0 \\
d_q \\
0 \\
\vdots \\
0
\end{array}
\right)=
\left(
\begin{array}{cc}
1 \\
0 \\
\vdots \\
0
\end{array}
\right)\in\mathbb{F}_2^{m+n}.
\end{equation*}

The condition $q>n$ implies $b_nc_p=0$, but $b_n=c_p=1$ since the polynomials $P_2(x)$ and $Q_1(x)$ have degrees $n$ and $p$ respectively. Therefore, it must hold $q\leqslant n$. Since $Q_2(x)$ has degree~$q$, then $d_q=1$ and
$
\left(1+b_{n-q}\right)=b_{n-q+1}=b_{n-q+2}=\ldots=b_{n-q+\min\{q,m-1\}}=0.
$
From $b_n=1$ it follows that $\min\{q,m-1\}<q$, that is $m\leqslant q$.
Finally, we get
$m\leqslant q\leqslant n<m
$
that is a contradiction.

Thus, in the case $m\leqslant n$ there exist invertible Sylvester matrices whose inverse are Sylvester matrices as well but for $m>n$ it does not hold.

\subsection{Problem ``Disjunct Matrices'' (unsolved)}

\subsubsection{Formulation}

Disjunct Matrices are used in some key distribution protocols for traitor tracing. Disjunct Matrices (DM) are a particular kind of binary matrices which have been applied to solve the Non-Adaptive Group Testing (NAGT) problem, where the task is to detect any configuration of $t$ defectives out of a population of $N$ items. Traditionally, the methods used to construct DM leverage on error-correcting codes and other related algebraic techniques.

Let $A = (x_1^\top, x_2^\top, \ldots, x_{N}^\top)$ be an $M\times N$ binary matrix. Then, $A$ is called $t$-\emph{disjunct} if, for all subsets of $t$ columns $S = \{x_{i_1},\ldots, x_{i_t}\}$, and for all remaining columns $x_j \notin S$, it holds that
\vspace{-0.2cm}
\begin{equation*}
\label{eq:t-disj} {\rm supp}(x_j) \not\subseteq \bigcup_{k=1}^t {\rm supp}(x_{i_k}),
\end{equation*}

\vspace{-0.2cm}
\noindent where ${\rm supp}(x)$ denotes the set of coordinate positions of a binary vector $x$ with 1s.

In other words, a matrix $A$ is $t$-disjunct if for every subset $S$ of $t$ columns the support of any other column is not contained in the union of the supports of the columns in $S$.

Prove what is the minimum number of rows in a $5$-disjunct matrix.

\subsubsection{Solution}

 We must admit that the formulation of the problem did not include the condition which makes this problem non-trivial. The condition is that the number of columns must be greater than the number of rows. This formulation comprises practical significance and has following equivalent form: given $t$, when does there exist a $t$-disjunct algorithm better than the trivial one that tests each item individually? Readers may find details regarding Non-Adaptive Group Testing (NAGT) problem together with known results and mentioned formulations in \cite{disjunct-16}.

The solution of the originally stated problem is $6$ --- consider the $6\times6$ identity matrix. This solution was discovered by several participants. However some participants (Alexey Chilikov from Bauman Moscow State Technical University, Aleksei Udovenko from University of Luxembourg, the team of Henning Seidler and Katja Stumpp from Technical University of Berlin) obtained bounds for the number of rows depending the parameter $t$ and the number of columns.

\section{Winners of the Olympiad}\label{winners}

\noindent Here we list information about the winners of NSUCRYPTO'2018 in Tables\;\ref{1-sc},\,\ref{1-st},\,\ref{1-pr},\,\ref{2-sc},\,\ref{2-st},\,\ref{2-pr}.

\renewcommand{\topfraction}{0}
\renewcommand{\textfraction}{0}

\begin{figure}[!h]
\centering
\includegraphics[width=0.95\textwidth]{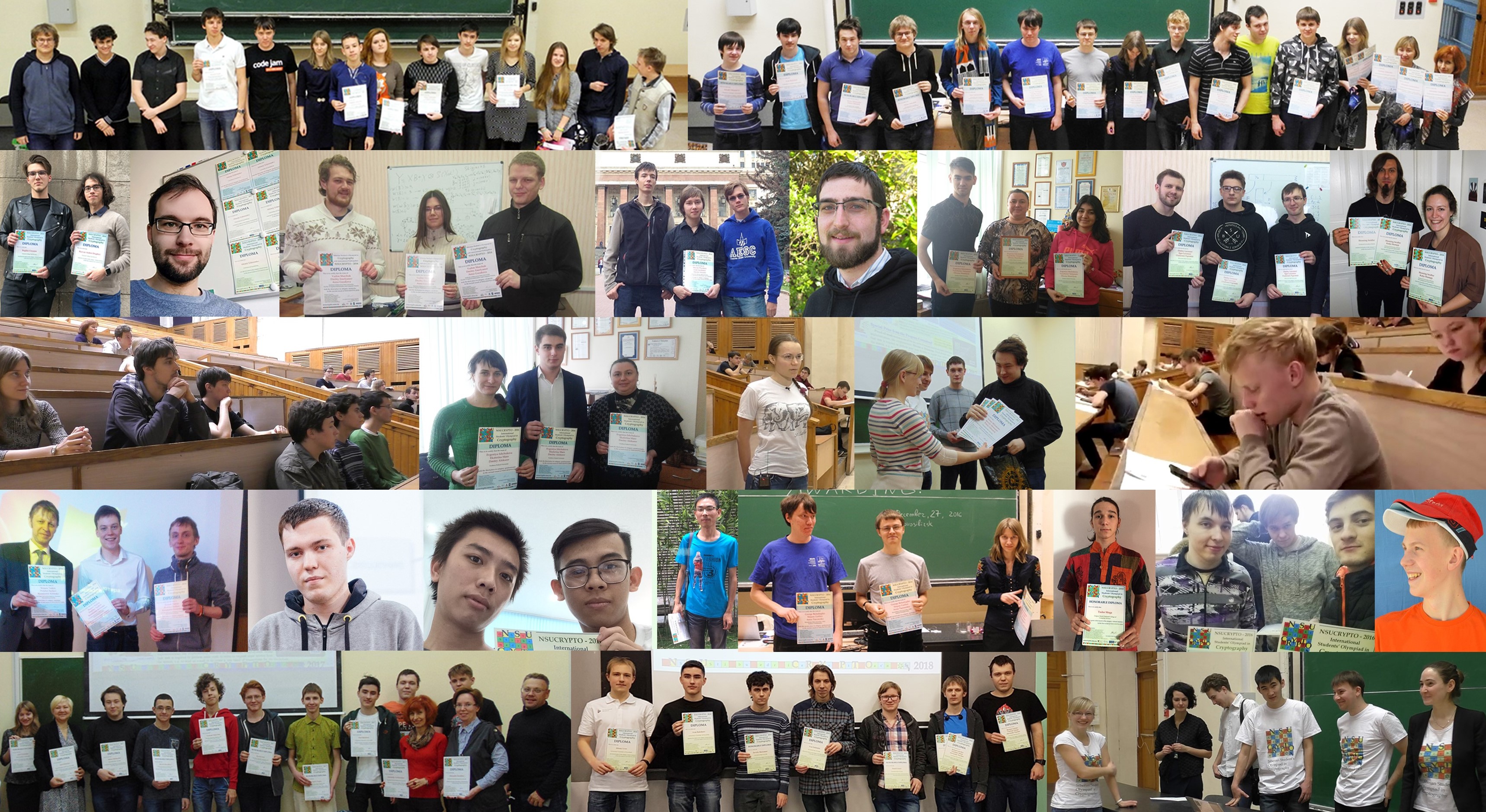}
\vspace{-3mm}
\caption{{\small Winners of NSUCRYPTO from 2014 to 2018}}
\label{fig:winners}
\end{figure}

\newpage
\newgeometry{top=1.0cm, bottom=0.5cm, left=2.6cm, right=2.6cm}
\thispagestyle{empty}

\begin{table}[!h]
\centering\footnotesize
\caption{{\bf Winners of the first round in school section A (``School Student'')}}
\label{1-sc}
\renewcommand{\arraystretch}{1.2}
\renewcommand{\tabcolsep}{1.6mm}
\medskip
\begin{tabular}{|c|l|p{2.9cm}|p{7.2cm}|c|}
  \hline
  Place & Name & Country, City & School &  Scores \\
  \hline
  \hline
  1 & Borislav Kirilov & Bulgaria,	Sofia &  The First Private Mathematical Gymnasium & 22 \\
   \hline
  1 & Alexey Lvov & Russia, Novosibirsk   & Gymnasium 6 & 21 \\
  \hline
  1 & Razvan Andrei Draghici &  Romania,	Craiova    & National College Fratii Buzesti  & 20 \\
  \hline
  2 &   Gorazd Dimitrov &    Macedonia,	Skopje & Yahya Kemal College &     18 \\
  \hline
  3 &   Ivan Baksheev	 &  Russia, Novosibirsk & Gymnasium 6 &     16 \\
  \hline
  3   &   Artem Ismagilov  & Russia,\newline 	Yekaterinburg    & The Specialized Educational and Scientific Center UrFU    &  16 \\
  \hline
  3   &   Bogdan Circeanu &  Romania,	Craiova &   National College Fratii Buzesti & 16 \\
  \hline
  3   &   Ruxandra Icleanu   & Romania, Craiova  &   Tudor Vianu National College of Computer Science  &   15\\
  \hline
  Diploma   &  Sofya Gorbunova	 &    Russia,	\newline Yekaterinburg &  The Specialized Educational and Scientific Center UrFU  & 13 \\
  \hline
  Diploma   &   Kirill Poltoradnev	 &  Russia,\newline 	Yekaterinburg & The Specialized Educational and Scientific Center UrFU  & 13 \\
  \hline
  Diploma   &   Tudor Moga	 &    Romania,	Brasov & Grigore Moisil National College of Computer \newline Science  & 11 \\
  \hline
  Diploma   &   Markas Cerniauskas &   Lithuania,	Kaunas    &  Kaunas Technology University Gymnasium  & 11 \\
  \hline
  Diploma   &  Mircea-Costin Preoteasa &  Romania,	Bucharest & Tudor Vianu National College of Computer Science & 11 \\
  \hline
  Diploma   &   Kirill Tugolukov &   Russia,	Ulan-Ude   & School 19, Olympiad training center ENTER  & 11 \\
  \hline
\end{tabular}
\end{table}

\vspace{-0.5cm}

\begin{table}[!h]
\centering\footnotesize
\caption{{\bf Winners of the first round, section B (in the category ``University Student'')}}
\label{1-st}
\renewcommand{\arraystretch}{1.2}
\renewcommand{\tabcolsep}{0.8mm}
\medskip
\begin{tabular}{|c|l|l|l|c|}
  \hline
  Place & Name  & Country, City & University & Scores \\
  \hline
  \hline
  1 &   Maxim Plushkin	& Russia,	Moscow &	Lomonosov Moscow State University    & 25 \\
  \hline
  2 & Robert Koprinkov &	Netherlands,	Nijmegen	& Radboud University &  20 \\
  \hline
  2 & Irina Slonkina	& Russia,	Moscow	    & National Research Nuclear University MEPhI  &    18\\
  \hline
  2 & Marc Houben	& Belgium,	Leuven	& KU Leuven & 18\\
  \hline
  3 & Dheeraj M Pai & India, Chennai    & Indian Institute of Technology Madras &  17\\
  \hline
  3 & Alexander Grebennikov	& Russia,	Saint Petersburg & Saint Petersburg State University & 16\\
  \hline
  3  & Roman Lebedev &	Russia,	Novosibirsk & 	Novosibirsk State University	 & 16 \\
  \hline
  3 & Dianthe Bose	& India,	Chennai & 	Chennai Mathematical Institute     &  15\\
  \hline
  3 & Ivan Sutormin	& Russia,	Novosibirsk	& Novosibirsk State University & 15\\
  \hline
  Diploma & Harikumar Krishnamurthy	& India,	Chennai  & Indian Institute of Technology Madras & 14\\
  \hline
  Diploma   & Roman Tarasov	& Russia,	Odintsovo &	Higher School of Economics &    13\\
  \hline
  Diploma & Sander Suverkropp &	Netherlands,	Wageningen	& Radboud University   & 13\\
  \hline
  Diploma   & Saeed Odak	& Iran,	Tehran & Khajeh Nasir Toosi University of Technology  & 13\\
  \hline
  Diploma & Thijs van Loenhout	& Netherlands,	Nijmegen &	Radboud University & 12\\
  \hline
  Diploma   & Daniil Gurev &	Russia,	Novosibirsk	& Novosibirsk State University	 &    12\\
  \hline
  Diploma & Neha Rino	& India,	Chennai &	 Chennai Mathematical Institute   & 12\\
  \hline
  Diploma   & Kristina Volyakova	& Russia, Yaroslavl	& Yaroslavl State University  & 12\\
  \hline
\end{tabular}
\end{table}

\vspace{-0.5cm}

\begin{table}[!h]
\centering\footnotesize
\caption{{\bf Winners of the first round, section B (in the category
``Professional'')}}
\label{1-pr}
\medskip
\renewcommand{\arraystretch}{1.2}
\renewcommand{\tabcolsep}{1.2mm}
\begin{tabular}{|c|l|l|l|c|}
\hline
Place & Name & Country, City & Organization & Scores \\
\hline
\hline
1 & Alexey Udovenko &   Luxembourg, Luxembourg  & University of Luxembourg &    23\\
\hline
3   &   Henning Seidler & Germany, Berlin   & TU Berlin &   16\\
\hline
Diploma & Alexey Chilikov	& Russia,	Moscow	  & Bauman Moscow State Technical University &  14\\
\hline
Diploma   & Samuel Tang	& Hong Kong,	Hong Kong	& Blocksquare Limited   & 12\\
\hline
Diploma &   Amedeo Sgueglia	& United Kingdom,	London & London School of Economics and Political Science	  & 12\\
\hline
Diploma &   Samad Alaamati	& Iran,	Tehran	& American Society for Industrial Security & 12\\
\hline
\end{tabular}
\end{table}

\begin{table}[h]
\centering\footnotesize
\caption{{\bf Winners of the second round (in the category ``School Student'')}}
\label{2-sc}
\renewcommand{\arraystretch}{1.2}
\renewcommand{\tabcolsep}{0.8mm}
\medskip
\begin{tabular}{|c|l|l|l|c|}
\hline
Place & Names  & Country, City & School & Scores \\
\hline
\hline
Diploma & Brian Ncube &	Zimbabwe,	Hwange &	Hwange High School  & 7\\
\hline
\end{tabular}
\end{table}

\vspace{-0.5cm}

\begin{table}[!h]
\centering\footnotesize
\caption{{\bf Winners of the second round (in the category
``University student'')}}
\label{2-st}
\renewcommand{\arraystretch}{1.2}
\renewcommand{\tabcolsep}{0.8mm}
\medskip
\begin{tabular}{|c|p{5.0cm}|p{2.8cm}|p{5.9cm}|c|}
\hline
Place & Name  & Country, City & University &  Scores \\
\hline
\hline
1 & Maxim Plushkin	& Russia,	Moscow &	Lomonosov Moscow State University  &       43 \\
\hline
2   & Irina Slonkina	& Russia,	Moscow	& National Research Nuclear University MEPhI   & 38 \\
\hline
2   & Dmitry Lavrenov, Egor Lavrenov, Uladzimir Paprotski &	Belarus, Minsk	& Belarusian State University    &   37\\
\hline
3 & Thanh Nguyen Van, Tuong Nguyen Van, Dinh Ton	& Vietnam,	Ho Chi Minh City & Ho Chi Minh City University of Technology, University of Science &  37\\
\hline
3   & Ngoc Ky Nguyen, Phuoc Nguyen Ho Minh, Danh Nam Tran	& Vietnam,	Ho Chi Minh City; France,	 Paris & Ho Chi Minh City Pedagogical University,	 Ho Chi Minh City University of Technology, Ecole Normale Superieure &      34\\
\hline
3 & Dianthe Bose, Neha Rino & India,	Chennai	  &  Chennai Mathematical Institute  & 33\\
\hline
3 & Roman Lebedev, Vladimir Sitnov, Alexander Tkachev	& Russia,	Novosibirsk & Novosibirsk State University & 30 \\
\hline
3  &   Mikhail Sorokin, Darya Frolova, Vladimir Bobrov	& Russia,	Moscow	&  National Research Nuclear University MEPhI  &   29\\
\hline
3  & Andrey Kalachev, Danil Cherepanov, Alexey Radaev &   Russia, Moscow  & Bauman Moscow State Technical University  & 27\\
\hline
Diploma & Saveliy Skresanov	& Russia,	Novosibirsk	& Novosibirsk State University &  21\\
\hline
Diploma & Harikumar Krishnamurthy, Aditya Pradeep, Dheeraj M Pai	& India, 	Chennai &   Indian Institute of Technology Madras     &  21\\
\hline
\end{tabular}
\end{table}

\vspace{-0.5cm}

\begin{table}[!h]
\centering\footnotesize
\caption{{\bf Winners of the second round (in the category ``Professional'')}}
\label{2-pr}
\renewcommand{\arraystretch}{1.2}
\renewcommand{\tabcolsep}{1.0mm}
\medskip
\begin{tabular}{|c|p{4.7cm}|p{2.8cm}|p{6.2cm}|c|}
\hline
Place & Names & Country, City & Organization  & Scores \\
\hline
\hline
1   &   Alexey Udovenko &   Luxembourg,\newline Luxembourg  & University of Luxembourg & 72\\
\hline
2   & Evgeniya Ishchukova, Vyacheslav Salmanov, Oksana Shamilyan	& Russia,	Taganrog & Southern Federal University & 46\\
\hline
3   &   Henning Seidler, Katja Stumpp & Germany, Berlin & Berlin Technical University &  40 \\
\hline
3   & Carl Londahl &    Sweden, Karlskrona & TrueSec AB & 38 \\
\hline

3   &Alexey Chilikov	& Russia,	Moscow	  & Bauman Moscow State Technical University
& 38\\
\hline

Diploma   &   Duc Tri Nguyen, Quan Doan, \newline Quoc Bao Nguyen &  Vietnam, \newline Ho Chi Minh city &CERG at George Mason University, \newline E-CQURITY, Ho Chi Minh City University of Technology&   37\\
\hline

Diploma &   Radu Caragea, Miruna Rosca, \newline Madalina Bolboceanu 	& Romania,	Bucharest &	 Bitdefender    & 32\\
\hline
Diploma &   Mikhail Polyakov, Mikhail \newline Tsvetkov, Victoria Vlasova	& Russia,	Moscow  & Bauman Moscow State Technical\newline University  & 30\\
\hline
Diploma &   Harry Lee, Samuel Tang  & Hong Kong, \newline Hong Kong & Blocksquare Limited, Hong Kong University of Science and Technology    & 29\\
\hline
Diploma & Lars Haulin	& Sweden,	Uppsala	& -- & 28\\
\hline
Diploma &  Sergey Titov, Kristina Geut  & Russia, \newline Yekaterinburg &  Ural State University of Railway\newline Transport &    22\\
\hline
Diploma &  Khai Hanh Tang, Neng Zeng, \newline Thu Hien Chu Thi &	Singapore, \newline	Singapore &  Ho Chi Minh City Pedagogical University, Nanyang Technological University &  21\\
\hline

\end{tabular}
\end{table}

\FloatBarrier

\newpage
\restoregeometry


\begin{thebibliography}{1}

\bibitem{nsucrypto-2015}
Agievich\;S., Gorodilova\;A., Idrisova\;V., Kolomeec\;N., Shushuev\;G., Tokareva\;N.
Mathematical problems of the second international student's Olympiad in cryptography~//
Cryptologia. 2017. V.\,41. No.\,6. P.\,534--565.

\bibitem{nsucrypto-2014}
Agievich~S., Gorodilova~A., Kolomeec~N., Nikova~S., Preneel~B.,
Rijmen~V., Shushuev~G., Tokareva~N., Vitkup~V.
Problems, solutions and experience of the first international student's
Olympiad in cryptography~//
Prikladnaya Diskretnaya Matematika (Applied Discrete Mathematics). 2015. No.\,3. P.\,41--62.

\bibitem{Agievich-2016}
Agievich~S., Marchuk~V., Maslau~A., Semenov~V. Bash-f: another LRX sponge function~// Cryptology ePrint Archive: Report 2016/587.

\bibitem{masking-1}
Bhasin~S., Carlet~C., Guilley~S. Theory of masking with codewords in
hardware: low-weight dth-order correlation-immune Boolean functions~// Cryptology ePrint Archive, Report 2013/303.

\bibitem{masking-5}
Bierbrauer J. Bounds on orthogonal arrays and resilient functions~// Journal of Combinatorial Designs. 1995. V.\,3. P.\,179--183.

\bibitem{masking-6}
 Friedman J. On the bit extraction problem~// Proc. 33rd IEEE Symposium on Foundations of Computer Science. 1992. P.\,314--319.

\bibitem{sharing}
Geut~K., Kirienko~K., Sadkov~P., Taskin~R., Titov~S.
On explicit constructions for solving the problem ``A secret sharing''~//
Prikladnaya Diskretnaya Matematika. Prilozhenie. 2017. No.\,10. P.\,68--70. (in Russian)

\bibitem{nsucrypto-2017}
Gorodilova A., Agievich S., Carlet C., Gorkunov E., Idrisova V., Kolomeec N., Kutsenko A., Nikova S., Oblaukhov A., Picek S., Preneel B., Rijmen V., Tokareva N. Problems and solutions of the Fourth International Students' Olympiad in Cryptography (NSUCRYPTO) // Cryptologia. 2019, Vol. 43, No. 2, pp. 138--174.

\bibitem{Obl18}
Oblaukhov~A. A lower bound on the size of the largest metrically regular subset of the Boolean cube~// Cryptography and Communications. 2018. Published online.

\bibitem{OA-2}
Picek S., Guilley S., Carlet C., Jakobovic D., Miller J.\,F. Evolutionary Approach for Finding Correlation Immune Boolean Functions of Order $t$ with Minimal Hamming Weight. In: Dediu AH., Magdalena L., Martn-Vide C. (eds) Theory and Practice of Natural Computing. Lecture Notes in Computer Science, vol 9477. Springer, Cham.

\bibitem{nsucrypto-2016}
Tokareva~N., Gorodilova~A., Agievich~S., Idrisova~V., Kolomeec~N., Kutsenko~A., Oblaukhov~A., Shushuev~G. Mathematical methods in solutions of the problems from the Third International Students' Olympiad in Cryptography~//
Prikladnaya Diskretnaya Matematika (Applied Discrete Mathematics). 2018. No.\,40. P.\,34--58.

\bibitem{disjunct-16}
 Shangguan~C., Ge~G. New bounds on the number of tests for disjunct matrices~// IEEE Trans. Inf. Theory. 2016, vol.\,62, no.\,12, pp.\,7518--7521.

\bibitem{FNV}
http://www.isthe.com/chongo/tech/comp/fnv/

\bibitem{moby-dick}
https://gist.githubusercontent.com/StevenClontz/4445774/raw/\newline1722a289b665d940495645a5eaaad4da8e3ad4c7/mobydick.txt

\bibitem{twinpeaks2}
https://nsucrypto.nsu.ru/olymp/2018/round/2/task/4

\end{thebibliography}
\end{document}